\documentclass[10pt]{article} 

\usepackage[utf8]{inputenc} 

\usepackage{geometry} 
\usepackage{graphicx} 
\usepackage{setspace}
\usepackage{amsmath, amssymb}

\usepackage{float} 

\usepackage{booktabs} 
\usepackage{array} 
\usepackage{paralist} 
\usepackage{verbatim} 
\usepackage{subfig} 

\usepackage{fancyhdr} 
\usepackage{sectsty}
\allsectionsfont{\sffamily\mdseries\upshape} 

\usepackage[nottoc,notlof,notlot]{tocbibind} 
\usepackage[titles,subfigure]{tocloft} 

\usepackage{cite}

\makeatletter
\renewcommand{\@biblabel}[1]{}
\makeatother

\begin{document}

\begin{center}
\Large{\textbf{Surface Tension Dominates\\Insect Flight on Fluid Interfaces}}\\
\end{center}

\noindent
\begin{center}
\textbf{Haripriya Mukundarajan$^1$, Thibaut C Bardon$^2$, Dong Hyun Kim$^1$ \& Manu Prakash$^{3\ast}$}\\
\end{center}
\noindent
$^{1}$ Department of Mechanical Engineering, Stanford University\\
$^{2}$ \'Ecole Polytechnique, Paris, 91128 Palaiseau, France\\
$^{3}$ Department of Bioengineering, Stanford University, Stanford, CA 94305, USA\\
Correspondence to $^{\ast}$manup@stanford.edu\\\\
\vspace{2 mm}

\noindent\textbf{RUNNING TITLE}\\
Interfacial insect flight - biomechanics\\\\


\doublespacing

\noindent\textbf{SUMMARY STATEMENT}\\
We present the first biomechanical model of insect flight on air-water fluid interfaces. We apply these insights to water-lily beetles, uncovering the complex interplay of aerodynamics, biomechanics and capillary forces.\\

\noindent\textbf{ABSTRACT}\\
Flight on the two-dimensional air-water interface, with body weight supported by surface tension, is a unique locomotion strategy well adapted for the environmental niche on the surface of water. Although previously described in phylogenetically basal aquatic insects like stoneflies, the biomechanics of interfacial flight has never been analyzed. Here, we report interfacial flight as an adapted behaviour in water-lily beetles (\textit{Galerucella nymphaeae, Linnaeus 1758}) which are also dexterous airborne fliers. We present the first quantitative biomechanical model of interfacial flight in insects, uncovering an intricate interplay of capillary, aerodynamic and neuromuscular forces. We show that water-lily beetles use their tarsal claws to attach themselves to the interface, via a fluid contact line pinned at the claw. We investigate the kinematics of interfacial flight trajectories using high-speed imaging and construct a mathematical model describing the flight dynamics. Our results show that nonlinear surface tension forces make interfacial flight energetically expensive compared to airborne flight at the relatively high speeds characteristic of water-lily beetles, and cause chaotic dynamics to arise naturally in these regimes. We identify the crucial roles of capillary-gravity wave drag and oscillatory surface tension forces which dominate interfacial flight, showing that the air-water interface presents a radically modified force landscape for flapping wing flight compared to air.

\newpage

\noindent\textbf{INTRODUCTION}\\
Insects constitute the majority of living species on land \cite{may}. Flight has played a crucial role in the proliferation of insects, enabling them to explore and adapt to new landscapes and habitats, evade predators and find mates \cite{simon}. About 98\% of insect species are capable of powered airborne flight \cite{simon} \textendash\space a phenomenon whose intricate biomechanics and origins has captivated physicists, engineers and biologists for centuries.\\ 

The quest to understand the evolution of flight has also led to studies on other mechanisms of locomotion in flightless insects that could have been precursors to powered flight. Some examples are ballooning on silk threads \cite{bell}, wafting on warm air currents \cite{washburn, chapman} and directed gliding during descent from tall trees \cite{dudleycatratus}. One hypothesis proposed for the origin of flight was surface skimming \textendash\space a phenomenon where wings or wing-like structures are used to generate propulsion along an air-water interface \cite{mardenthomas, marden}. Several aquatic species of stoneflies and mayflies use their wings to row, sail, or flap along the air-water interface. Some of these have rudimentary wings and cannot generate sufficient lift to completely support their body weight in air, whereas others use surface skimming in a context-dependent fashion which often comprises cold temperatures not permissive for flight. A competing evolutionary hypothesis poses directed gliding as the origin of insect flight \cite{dudley1, dudley2}, based on observations in creatures such as the canopy ant \textit{Cephalotus atratus}. In any event, flight along a fluid interface is a fascinating mode of insect locomotion that is far less studied than other modes of aerial or interfacial locomotion. The relative merits of the interfacial flight origin hypothesis have been discussed by comparing observational, molecular and fossil evidence, and lie outside the scope of our current study. Here we present a quantitative analysis of the flight dynamics involved \textendash\space a missing piece of the puzzle required for a more complete understanding of interfacial flight. In this work, we present the first biomechanical model of the physics underlying interfacial flight, to elucidate the roles of capillary, aerodynamic and neuromuscular forces in giving rise to this rare and complex behaviour.\\

We define interfacial flight as biomechanically powered flapping-wing locomotion where the insect’s trajectory is limited to the two-dimensional plane of the air-water interface. Hence the spatial degrees of freedom that need to be controlled by the insect are reduced from six to just three. Although the insect's body weight is now supported by surface tension, several complications due to interfacial forces quickly arise. Firstly, water is fifty times more viscous than air. In addition, it has a large surface tension which can exert capillary forces as high as a few hundred times the viscous drag during interfacial flight (the ratio of capillary to viscous forces is given by the inverse of the capillary number, $Ca^{-1}=\frac{\sigma}{\mu V}$, which is greater than $100$ for all characteristic velocities in water below about $0.7$ ms$^{-1}$.). Further, the insect generates capillary waves or ripples when moving along the interface, which produce non-linear drag effects depending on the speed of motion. Finally, the forces exerted by surface tension that keep an insect attached to the interface are dependent on the wetting properties of its body parts in contact with water. This resultant force of surface tension can be comparable to or greater than the body weight. It can either be directed upwards (supporting an insect's body weight) as commonly seen in water-walking insects, or directed downwards (trapping an insect on the water surface) as commonly seen in insects that are trapped on a water surface and unable to take off. Overall, the non-linear effects of capillary forces, wetting of different parts of the insect's body and the complexity of contact line dynamics on insect body parts during the transition from interfacial to airborne flight are not well understood. The phenomenon of interfacial flight and the transition to airborne flight thus presents intriguing puzzles for studies in physical biology.\\

Here, we report the observation of two-dimensional interfacial flight as a derived adaptation of three-dimensional aerial flight in waterlily beetles (\textit{Galerucella nymphaeae}), which are capable of interfacial flight over a wide variety of speeds ranging up to a rapid 0.5 ms$^{-1}$, and can also make smooth transitions between both interfacial and airborne flight modes. This makes the waterlily beetle an ideal organism to analyze the fundamental physics of interfacial flight and the stepwise transitions to airborne flight. Using high speed videography, we recorded the kinematics of flight trajectories on the interface and identified features that are strongly influenced by capillary forces. We used our observations to develop a quantitative mathematical framework for modelling naturally confined flapping wing propulsion on a two-dimensional fluid interface. Our results show that as locomotion velocities horizontally along the interface increase, interfacial flight becomes energetically more expensive due to the influence of capillary wave drag. In addition, a chaotic regime comes into existence as vertical lift forces increase at a given wing flapping frequency. These insights highlight that two-dimensional interfacial flight is a complex non-linear phenomenon whose unique kinematics differs significantly from purely airborne flight.\\\\

\noindent\textbf{MATERIALS AND METHODS}\\
About $400$ live adult beetles (\textit{G.nymphaeae} and \textit{G.pusilla}) were captured in Harvard Forest, Massachusetts and in Montana during the summer months of May-July over several years. These were maintained in the lab in an incubator at about $18^{\circ}$C and 60\% RH, with a 14-hour daylight cycle. About 50 beetles at a time were kept in ($300$ mm)$^3$ cages with a water-filled $150$ mm Petri dish placed inside, and fed water-lily or loosestrife leaves collected at the capture site. Beetles were carefully transferred using a piece of water-lily leaf, to avoid touching them or accidentally wetting their legs prior to observation. One beetle at a time was placed in a smaller cage ($200$ mm X $150$ mm X $100$ mm) with about an inch of standing water, and bits of leaves or green plastic pieces placed on the opposite end of the cage to initiate motion. All experiments were carried out in laboratory conditions where the temperature was maintained around $23^{\circ}$, so as not to vary the surface tension of water. \\
 For kinematic measurements, we took high speed videos of interfacial flight at $3000$ frames per second using a Phantom v1210 camera. Wide field videos with typical field of view around $250$ mm but low resolution on the order of $100\thickspace\mu$m per pixel were used to observe full flight trajectories (Movie SV3, SV4, SV7, SV8, SV10, SV11). High magnification videos with smaller fields of view around $100$ mm but high resolution on the order of $10\thickspace\mu$m per pixel were recorded to measure displacements and velocities for parts of some of the trajectories (Movies SV1, SV2, SV9, SV12). Certain points on the beetle's body, such as the eye, the mouthparts and the femur-tibia joint, stand out as high-contrast areas that can be used as natural markers. We used an in-house shape-tracking script to automate the extraction of coordinates of the centroids of these markers in each frame. Although a significant number of video runs were collected, only trajectories with sufficient pixel resolution that were perfectly parallel to the imaging plane were chosen for quantitative analysis.\\
Simulations of the dynamics were performed in MATLAB R2012b using a built-in ODE solver (ode113), with relative tolerances of $10^{-9}$ and constant integration time step of $10^{-8}$ seconds. The elimination of numerical noise was verified by computing take-off times for a variety of different thrust-to-lift ratios $p$ keeping other parameters constant, and checking that take-off time is repeatably constant for a given lift $L$ at the required precision, as expected from decoupling of horizontal and vertical dynamics.\\\\
 
\noindent\textbf{RESULTS}\\
\textbf{Discovery of interfacial flight in wild \textit{G.nymphaeae}}\\
We report the discovery of flapping wing interfacial flight in beetles of the genus \textit{Galerucella} (Figs 1A, 1B, Movies SV1, SV2). Unlike many other interfacial fliers, all species of \textit{Galerucella} possess well-developed airborne flight capabilities (Movies SV3, SV4, SV5). However, some species like the waterlily beetle \textit{G.nymphaeae} display a preference for interfacial flight, while others like the purple loosestrife beetle \textit{G.calmariensis} are incapable of it. In the wild, waterlily beetles (\textit{G.nymphaeae}, formerly \textit{Pyrrhalta nymphaeae}) live on the surface of ponds in the temperate forests of North America where they feed on floating waterlily leaves. They react to visual cues of floating green objects, executing straight line movements along the surface towards these objects at high speeds of over fifty body lengths per second ($0.3$ ms$^{-1}$). Here we report that \textit{G.nymphaeae} are among the fastest insects on the air-water interface \cite{bush, mardenthomas}. The preference for interfacial flight is pronounced under a variety of conditions, with the beetles skimming along the interface between leaves at a variety of different temperatures between $14^{\circ}$C and $28^{\circ}$C. This is the range of temperatures at which we observed the beetles, ranging from ambient environmental conditions to laboratory conditions. Since two-dimensional flight behavior is observed in waterlily beetles when traversing the pond surface from one leaf to another, we conjecture that topologically, interfacial flight provides a more efficient foraging mechanism to feed on floating waterlily leaves on the planar water surface using motility that is also confined to two dimensions.\\

\noindent\textbf{Laboratory observations of interfacial flight}\\
After preliminary observations in the field, we brought the waterlily beetles into the lab and recorded observations of interfacial flight in the lab environment. \textit{G.nymphaeae} prepares for flight along the interface by first lifting each leg off the surface and setting it back down. This ensures that there is no excessive wetting of the tarsi due to impact or other accidental contact, such that water enters the gaps between the hydrophobic hairs on the leg and increases contact area, though fouling due to naturally occurring floating lipids in wild settings is not so readily overcome. Just before the start of interfacial flight, the beetle lifts its two middle pair of legs off the water and angles its body upwards (Movie SV1). This posture with only four legs in contact with water lowers the drag from the legs, but still provides a wide base to prevent falling over during flight. Such a posture also serves to place the legs out of the way of the stroke of the dominant hindwing. This is a significant difference in posture compared to stoneflies and mayflies, which have dominant forewings and hence raise their forelegs up during skimming \cite{mardenthomas}. The beetle then flaps its wings to unfurl them fully, and sets up consistent, strong wingbeats resulting in horizontal speeds of up to $0.5$ ms$^{-1}$. This is not only fast relative to its body length of $6$ mm, but in absolute terms, to the best of our knowledge, is among the fastest reported average horizontal speeds for an insect on a fluid interface \cite{bush, mardenthomas}. The flapping of well-developed wings allows both force generation and high speeds to be continuously sustained over relatively long distances of a few meters (Fig. 1E,F). Though the insect itself is difficult to see during its quick movement, a striking visible manifestation of its motion is the continuous formation of a train of capillary waves (ripples) around it (Fig. 1C, Movie SV6). Such waves have also been reported in other surface-skimming stoneflies \cite{mardenthomas}. They have also been shown to play an important role in the movement and behaviour of insects like whirligig beetles, which are partially submerged in water \cite{casasvoise2010}. This insight highlights the importance of considering capillary dissipation effects in interfacial flight analyses.\\

\noindent\textbf{Structural adaptations enabling 2D flight}\\
Using electron microscopy, we discovered that the waterlily beetle's body and legs are covered in small, water-repellent setae that can form a plastron air bubble (Fig. 2B,D,E), thus making the entire body superhydrophobic. This observation is intutive, since though the insect lives on the air-water interface, it does not become wet during its fast spurts of movement on the water surface. More surprisingly, we find that the water repellent tarsi of the legs end in a smooth hydrophilic pair of curved tarsal claws, which are submerged into water. We hypothesized that the claws are hydrophilic and used to directly anchor the insect to the interface during flight (Fig. 2A,C,J-L). This difference in wetting properties between the tarsi and claws forms a very distinctive hydrophobic-hydrophilic line junction on the insect’s legs. Many aquatic insects have entirely superhydrophobic leg surfaces that support them on water without any submerged parts \cite{prakash}. However, the presence of a boundary between non-wetting tarsi and wetting claws on \textit{G.nymphaeae} instead pins the water meniscus at this boundary, with the tarsi in air and claws in water (Figs 1B inset, 2K-L, Movies SV2, SV12). Here, the term pinning indicates that the contact line does not move relative to the leg of the beetle, and always remains at the boundary between the two structures with different wetting properties. Such differential wetting stabilizes the meniscus at this line junction between the claws and tarsi \cite{barthlott}, preventing the contact line from slipping up and down the claws. This ensures the continuous action of capillary forces at the tarsus-claw joint, allowing capillary effects to dramatically influence interfacial flight. The contact line pinning at the claw joint thus tethers adult \textit{G.nymphaeae} to the interface during flight, as surface tension now counteracts not only the downward pull due to weight during wing upstroke, but also the upward pull of wing forces during the downstroke. We estimate the maximum restoring force to be equal to $4*2\pi\sigma R_{claw}\approx 5*mg\approx 100\thinspace\mu$N, where $\sigma=0.072$ Nms$^{-1}$ is the surface tension of water, $R_{claw}=57\thinspace\mu$m is the radius of the contact line at the tarsus-claw joint, and $m=2.2$ mg is the mass of the beetle. (The multiplication factor of 4 refers to the number of legs simultaneously engaged in normal two-dimensional flight in waterlily beetles.) The insect's body weight is supported by an opposite deflection of the meniscus which prevents sinking. This can be estimated by the non-dimensional Bond number (defined as the ratio of weight to surface tension, Bond number $Bo_{y}\approx 0.2$). Thus, the force exerted by surface tension at the claw joint opposes both upward and downward motions of the beetle, constraining its flight to an effectively planar fluid interface. It is interesting to note that the larvae of \textit{G.nymphaeae} have also been previously reported to have adapted to climbing fluid menisci near lily leaf edges by deforming the air-water fluid interface at the points of immersion of their wetting claws \cite{bushhu}.\\

\noindent\textbf{Kinematics of interfacial flight trajectories}\\
We first quantify the kinematics of flight on the two-dimensional interface. The motion at the point of immersion of the beetle's legs is of maximum interest, since the effects of surface tension act directly to modify motion at this point. To extract immersion point trajectories, we tracked the femur-tibia joint which is the closest natural high-contrast marker on the hind legs. This joint is expected to follow the immersion point closely as there is no observed rotation of the intervening tarsal joints, hence it was preferred instead of the centre of mass of the beetle. One representative interfacial flight trajectory is shown in Fig. 3A, with a few additional trajectories in Fig. S1C,D for comparison. Trajectories recorded using markers other than the femur-tibia joint are shown in Fig. S1E,F, and have joint rotations and postural changes superimposed on the motion.\\

We observe several differences in the kinematics between interfacial and airborne flight in \textit{G.nymphaeae}. We identified five important kinematic trends at the femur-tibia joint, which are illustrated in the representative trajectory in Fig. 3A, extracted from Movie SV1. These trends are also observed in other trajectories, and are characteristic of interfacial flight in \textit{G.nymphaeae}. First, the wingbeat frequency remains constant after the initial stages of flight initiation, with frequency measured at $116\pm5$ Hz (Fig. S1A). Second, we observe a sigmoidal variation in horizontal displacement in each wingbeat, with the steepest increase occurring during the downstroke (Fig. 3B). Third, average horizontal velocity along the interface increases to values reaching up to $0.5$ ms$^{-1}$, before leveling off at some terminal velocity close to this value. Within each wingbeat however, velocity varies semi-sinusoidally, reaching its maximum value during  downstroke and taking a sharp dip during upstroke (Fig. 3C). Fourth, at higher resolutions, we observed that the trajectory displays striking vertical oscillations perpendicular to the interfacial plane, which are often slightly phase shifted relative to the wingbeat. Such a shift in relative phase results from the interaction of a restoring surface tension force with oscillatory forces produced by the wings, and provides direct evidence of capillary effects on the trajectory.
 This vibration of the trajectory is the most intriguing feature of interfacial flight, and arises from the restoring force exerted by surface tension against the wing lift. The oscillations have a frequency close to the wingbeat and amplitude spanning between $300\thinspace\mu$m and $500\thinspace\mu$m, or up to 25\% of the insect's dorso-ventral height of $2$ mm. The maximum peak-to-peak displacement amplitudes are approximately equal to twice the theoretical maximum meniscus height that can be supported by gravity. (Assuming quasi-static zero-pressure deformation of the meniscus, twice the maximum height $2*H_{max}=2*R_{claw}\ln(\frac{2\kappa^{-1}}{R_{claw}})\approx500\thinspace\mu$m where $R_{claw}=57\thinspace\mu$m is the radius at the claw and $\kappa^{-1}=2.71$ mm is the capillary length.) This length scale is a fundamental characteristic of interfacial flight trajectories, and any further increase in oscillation amplitude would result in meniscus breakage and airborne flight. Thus, the amplitude of vertical oscillations in the trajectory determines the transition between interfacial and airborne flight. Accelerations due to the vertical oscillations are of the order of g, with variations of similar magnitude due to the relative phase between the wingbeat and the oscillations (data not shown). This shows that the oscillations contribute significantly to the dynamics of the trajectory with storage and release of energy from the water meniscus similar to a fluid trampoline \cite{giletbush}. Lastly, the peak displacement of each oscillation varies widely with each successive wingbeat, appearing as though there is an additional irregular variation superimposed on the oscillations at wingbeat frequency. We thus infer that the interactions between surface tension and wingbeat modify the oscillations from simple sinusoids to more complex trajectories, as the relationship between surface tension and vertical displacement is not a simple linear proportional dependence but a non-linear hyperbolic function instead. Taken together, these trends highlight that interfacial flight in \textit{G.nymphaeae} is extremely fast due to its well-developed flapping wing flight capabilities, and that non-linear oscillatory effects caused by surface tension modify interfacial flight trajectories significantly compared to airborne flight.\\

\noindent\textbf{Force landscape during interfacial flight}\\
The kinematic data in Fig. 3A allows us to estimate the relative importance of the forces involved in interfacial flight using nondimensional parameters (see Supplementary Information for detailed calculations). To compare the importance of inertial and capillary forces in both horizontal and vertical translation, we calculate the well known non-dimensional Weber number, defined as the ratio of inertial to capillary forces. $We=\rho \dot{x}^2 R_{l}/\sigma$ where $\rho$ is the density of water, $\dot{x}$ is the translational fluid velocity and $R_{claw}$ is the characteristic length scale \textemdash\space the leg radius at the claw joint. For both horizontal and vertical motion, we obtained a fairly low Weber number on the order of magnitude of 0.1 ($We\sim\mathcal{O}(0.1)$), implying that capillary forces dominate. Next we look at the Reynolds number, defined as the ratio of inertial to viscous forces. $Re= \dot{x} R_{claw}/\nu$, where $\nu$ is the kinematic viscosity of the fluid. In both water and air, we find moderate values of $Re\sim\mathcal{O}(10-100)$. We also compute the Strouhal number in water, defined as the ratio of added mass of water at the legs in each wingbeat to inertial forces, as a small value $St=\frac{fR_{claw}}{\dot{x}}\approx0.02$ indicating that added mass of water can safely be ignored. From this, we conclude that interfacial flight is dominated by capillary forces with smaller contributions from water inertia and aerodynamic forces, while viscous dissipation is small. To corroborate the relative dominance of capillary forces over viscous forces for horizontal motion, we show relatively low values of the Capillary number. The Capillary number, defined as the ratio of viscous to capillary forces, is found to be $Ca=\frac{\mu\dot{x}}{\sigma}\approx0.005$ \textendash a small value indicating that viscous forces are small in comparison to capillary forces. The prominent role of capillary forces makes the physics of interfacial flight unique in comparison to both water walking arthropods where capillary forces are secondary to viscous drag forces \cite{bush, suter}, and also airborne flight where capillary effects are absent and aerodynamic forces prevail \cite{sane, dickinsonlehmannsane, lehmann, ellington}. Next we use these insights to construct a reduced order model for interfacial flight, which captures the essential physical phenomena involved.\\

\noindent\textbf{Dynamic model for interfacial flight}\\
Here we construct the simplest possible model of interfacial flight taking into account interfacial, viscous, gravitational and aerodynamic forces (Fig. 4). The dynamical analysis of interfacial flight can be greatly simplified by reducing the insect to a single particle pinned at the air-water interface with all forces acting directly on it, making the problem analytically tractable. We outline the main conceptual elements of our model here, with a detailed treatment of the mathematical equations and their validation provided in the Supplementary Information. The interfacial contact is represented by the two forces of capillary-gravity wave drag $C_{x}$ in the horizontal direction $X$ along the interface \cite{degennesraphael} and the resultant of surface tension $S_{y}$ in the vertical direction $Y$ normal to the interface. Capillary-gravity wave drag $C_{x}$ is the dissipative force that arises when the insect moves at speeds sufficiently high to exceed the minimum phase velocity of capillary-gravity waves on the interface, causing its momentum to be radiated away by the waves. Finally, the force of surface tension $S_{y}$ acts along the vertical direction, both supporting the insect's body weight and providing a restraint against pulling off the surface depending on the nature of the meniscus curvature. There is considerable variation in the geometry of contact for the legs of different insects, such as line contact, contact through groups of penetrating hairs \cite{prakash} and also through penetrating hydrophilic ungui \cite{bushhuprakash}. Adhesion forces arising from surface tension must be modelled taking the particular nature of the contact into account. Here, in the case of \textit{G.nymphaeae}, we use an adhesion model corresponding to pinning of the meniscus at the line boundary between the superhydrophobic leg and the hydrophilic claw. Surface tension $S_{y}$ is assumed to arise from the quasi-static deformation of non-interacting minimal surface menisci at each leg, which allows us to simplify the force model by ignoring contact angle hysteresis and interactions between the menisci. These approximations are justified as the capillary relaxation time scale ($\approx 50\thinspace\mu$s) is much smaller than the inertial timescale (wingbeat period $\tau=8.67$ ms), meaning that any flows within the meniscus or contact line hysteresis die out fast enough that they can be ignored and meniscus deformations assumed to be instantaneous. Also, the meniscus maximum height ($\approx 250\thinspace\mu$m) is much smaller than the capillary length $\kappa^{-1}=2.709$ mm, which in turn is smaller than the physical separation between the legs corresponding to the beetle's body width and length of $4$ to $6$ mm, indicating that the menisci are placed far enough apart that their finite physical size is too small for them to interact. The minimal surface assumption is verified by dipping the insect's claw into water and fitting a minimal surface profile to the image of the meniscus (Fig. 2K). The meniscus produces a resultant force of surface tension that is opposite the direction of vertical displacement with a non-linear dependence on the vertical displacement \cite{degennes}, like a non-linear spring. The forces common to both interfacial and airborne flight are the wing forces (horizontal wing thrust $T_{x}$ and vertical wing lift $L_{y}$), air drag $A_{x}$,  and gravity $G_{y}$. Since the submerged claw of the legs dissipate some energy in water, we also add the small drags $W_{x}$ and $W_{y}$ from the water bulk. By resolving forces along horizontal and vertical directions, our model describes the dynamics of interfacial flight by the two decoupled scalar equations in $X$ and $Y$ \textendash\space \begin{equation}m\ddot{x}=T_{x}-W_{x}-A_{x}-C_{x}\end{equation} \begin{equation}m\ddot{y}=L_{y}-W_{y}-G_{y}-S_{y}\end{equation}
This model of an insect in interfacial flight is equivalent to a particle pinned to the plane of the fluid interface and translating in-plane, while also executing forced damped non-linear oscillations out-of-plane (Fig. 4). By simulating a variety of different trajectories and comparing them with experimental data, our model allows us to gain additional quantitative insight into the dynamics of interfacial flight. This model is broadly applicable to all modes of interfacial flight, and its individual force terms can be modified to represent various cases such as high-speed flight, wetting of legs or varying wing force as appropriate for different insects. Here, we apply our model to interfacial flight in \textit{G.nymphaeae} and present our insights into the energetics and dynamics of high-speed interfacial flight and takeoff into air.\\

\noindent\textbf{Body angle can be varied to produce different modes of interfacial, airborne or backward flight}\\
A correlation between increasing body angle and the progression from interfacial to airborne flight was first proposed when comparing the different modes of interfacial flight in stoneflies \cite{mardenkramer, mardenthomas}. Since \textit{G.nymphaeae} displays more than one of these flight modes as well as airborne flight, this makes it an ideal organism to study the effect of postural changes on transitions between modes. We observed in our experiments that take-off into air always begins with a postural change, usually a transition from 4-leg to 2-leg interfacial flight, before culminating in airborne flight (Movies SV3, SV4, SV5, SV7, SV12). In addition, we also observe many instances of backward flight where the insect's horizontal velocity is directed opposite to its dorso-ventral axis (Movie SV8). Based on these observations, we hypothesized that postural changes, particularly related to the stroke plane and body angles (defined in Fig. 1D), are key to adopting different flight modes. Here we define four distinct flight modes \textemdash\space interfacial 4-leg, interfacial 2-leg, airborne forward and airborne backward. We analyzed videos of 12 different flight sequences containing one or more of these modes in each, and measured stroke plane and body angles of the insect in each wingbeat. In all four flight modes, the angle between the dorso-ventral axis and the stroke plane, corresponding to wing joint rotation in the sagittal plane, is confined to a narrow range spanning only $30^{\circ}$ (Fig. S1B). This implies that a given body angle for the insect allows only a restricted range of stroke plane angles. Indeed, Fig. 5A shows that as body angle of the insect increases, the stroke plane angle correspondingly decreases in a strongly linear correlation. Further, with increasing body angle, the flight mode changes from interfacial to airborne to backward. This is made evident by the distinct clustering of wingbeat data from different flight modes at different regions on the best fit correlation line. Thus, the transition from interfacial to airborne to backward flight is characterized by increased body angle and decreased stroke plane angle (Fig. 5B). We hence propose that the insect transitions between flight modes by altering its stroke plane angle within the permissible joint rotation range, resulting in a torque on the body that changes the body angle. This change in body angle can be reduced, maintained or further increased through a feedback mechanism, where the stroke plane angle is respectively increased, kept constant or decreased again. We further hypothesize that this change in posture causes the variation in flight mode by altering the distribution between lift and thrust for the same net force exerted by the wings. Greater thrust at lower body angle corresponds to interfacial flight at high speeds, while greater lift at higher body angle corresponds to efficient vertical airborne flight. This is corroborated by using our model to run computer simulations of flight sequences at constant wing force and varying thrust-to-lift ratio $p$ (Fig. 6B), where a series of trajectories smoothly varies from interfacial to airborne to backward flight. Thus, we show that body angle is a critical parameter that can be tuned by an insect, altering the distribution of wing force between lift and thrust in order to achieve transitions between different modes of interfacial and airborne flight.\\

\noindent\textbf{Wing lift can be estimated using surface tension}\\
In the majority of instances of interfacial flight in \textit{G.nymphaeae}, we observe successful transitions to airborne flight (Fig. 2L, Movies SV3, SV4, SV5, SV7, SV12). However, in some rare cases, the take-off attempts naturally fail (Fig. 5C, Movie SV9) due to impacts, accidents or fouling by floating organic compounds, which cause water to be driven into the gaps between its leg hairs displacing the air initially trapped within. This corresponds to a change in the wetting state of the legs, from the hydrophobic Cassie state where an air layer reduced the water contact area, to the more hydrophilic Wenzel state where the air is displaced and the contact line perimeter drastically increases \cite{bushhuprakash}. In the failed attempt shown in Movie SV9, the insect transitions from 4-leg to 2-leg contact with the interface, attempting to concentrate its entire wing force in the vertical direction for taking off. The body is angled almost vertically, with the wings beating in the horizontal plane (Fig. 5D). Surprisingly, the insect comes to an absolute halt in the horizontal direction as it now produces only lift and no thrust via its wing beats (Fig. 5E). In this current Wenzel state of wetting at the legs, surface tension no longer discontinuously drops to zero at this limiting height, but maintains its maximum value since the leg now behaves like a wet fibre being pulled out of a fluid bath. Though such a situation is momentarily encountered during takeoff, it is continuously maintained for some time in this special case. In such a situation, the beetle is almost stationary in horizontal and vertical directions, and lift forces are exactly balanced by surface tension and body weight. This corresponds to the 1D static equilibrium case where $\ddot{y}, \dot{y}, \ddot{x}, \dot{x}, T_{x}=0$. The model hence reduces to \begin{equation}\label{stall}L_{y}=S_{y}+G_{y}\end{equation} This force equilibrium can be used to indirectly measure instantaneous wing forces by measuring the meniscus height, which is equal to the vertical displacement of the claw joint. Forces measured under these special conditions help to establish upper bounds for the typical lift forces exerted by an insect in normal interfacial flight, and provide a good estimate for the forces required to takeoff directly from the interface into air in the absence of oscillations. The successful takeoff attempt in Movie SV12 clearly shows the lifting of a meniscus to its maximum height and subsequent breaking at the instant of takeoff. In the failed attempt shown in Movie SV9, we also observe that the meniscus is lifted to its maximum height, and surface tension assumes its maximum value. We estimated the corresponding maximum lift force as $L_{y}=2\pi\sigma R_{claw}+mg\approx73\thinspace\mu$N. ($R_{claw}=57\thinspace\mu$m is the leg radius at the immersion point measured from the scanning electron micrograph in Fig. 2A, and $m=2.2$ mg is the mass of the insect, found by averaging values measured for 30 recently deceased insects). This magnitude of lift force corresponds to a lift coefficient of $C_{L}\approx1.55$. This $C_{L}$ is slightly higher than the typical $C_{L}$ values of $1.1$ to $1.4$ measured experimentally for live tethered insects \cite{dudley,dickinsonlehmannsane}, or found by computational methods \cite{suntang}. Further, the lift-to-weight ratio in this experiment is $q=3.4$. This value of $q$ is significantly higher than the typical values of $1$ to $2.5$ observed in airborne flight \cite{dudley}, though it has been shown to be achievable in measurements made on insects with high flight muscle ratios carrying maximal weights \cite{marden1987}. (For detailed calculations of $L_{y}$, $C_{L}$ and $q$, please see the Supplementary Information.) In the case where all $4$ legs are in contact with the meniscus, the maximum force required for takeoff is even higher, corresponding to $q=5.8$. These increased $C_{L}$ and $q$ values indicate that for an insect without adaptations enabling the characteristic kinematics and force landscape of interfacial flight, making the transition from interfacial to airborne flight through purely quasi-static means would need much greater lift forces compared to typical airborne flight, primarily to escape the strong pull of surface tension. This challenges the limits of flight performance and necessitates high flight muscle ratios. Thus the interfacial energy landscape can also be thought of as an energy trap, which in the absence of appropriate adaptations requires a very high threshold energy to break the meniscus and take off into air.\\

\noindent\textbf{Interfacial flight has higher total drag forces than airborne flight at typical speeds}\\
The total drag force acting upon an insect in interfacial flight is the sum of the fluid resistance to motion arising from three distinct sources \textendash aerodynamic, hydrodynamic and capillary wave drag. An insect experiences comparable aerodynamic drags in interfacial and airborne flight, as the body and wing area normal to the air flow is about the same. However, an insect in interfacial flight has to contend not only with this aerodynamic drag, but also the additional horizontal resistances from capillary-gravity wave drag and hydrodynamic drag. Using our model, we computed the drag at different velocities for \textit{G.nymphaeae} in airborne flight and interfacial flight (Fig. 6A) \cite{chepelianskii}. Capillary-gravity wave drag is absent at speeds below the threshold of $c_{min}=0.23$ ms$^{-1}$, but quickly exceeds air drag slightly above this critical speed. At the typical speeds of $0.3$ to $0.5$ ms$^{-1}$ observed in interfacial flight in \textit{G.nymphaeae}, capillary-gravity wave drag is several times higher than aerodynamic drag. This means that it is much easier for an insect to fly in air at these speeds than along the interface. Total drag forces in interfacial and airborne flight do converge to comparable values at speeds above $\sim 3\thinspace c_{min}$, as Weber numbers increase and the relative dominance of surface tension decreases. Different models come into play at these high Weber number regimes to give more accurate estimations of the capillary-gravity wave drag on three-dimensional submerged hulls \cite{sunkeller}. However, this is well above the terminal velocities observed for interfacial flight, passing into the regime where drag forces are comparable to or exceed the wing forces. Hence, interfacial flight is considerably more energetically expensive than airborne flight except either at very low or very high velocity. This simple calculation rules out energetic advantages for two-dimensional flight in G.nymphaeae, primarily hinting towards its relevance either as a better search strategy in the two-dimensional landscape on the water surface, or as a viable low speed alternative to self-supported airborne flight at colder ambient temperatures where flight muscles are less efficient.\\

\noindent\textbf{Vertical oscillations play an important role in assisting takeoff from the interface}\\
Here, we discuss the dynamics by which an insect in interfacial flight can meet the high energy requirements to take off into the air. Our observations of failed takeoff attempts underline that surface tension is a powerful energy trap, which makes the net force required to pull off the interface higher than that needed to stay aloft in air. This is also evident in many modern flying insects, such as flies and mosquitoes, which become trapped on a water surface when their limbs are wetted. We applied our dynamic model to understand how \textit{G.nymphaeae} is consistently able to take off with ease using wing forces alone. Vertical oscillations are a feature of interfacial flight that is both observed in experimental trajectories (Fig. 3A) and predicted by our computer simulations (Fig. 6B). Though the lift required to break off the interface in the static case is high, it is reduced for a dynamic flight sequence because of the significant contribution from inertia due to the vertical oscillations. When the insect is in motion, the acceleration of the beetle produced by the oscillations is of the same order of magnitude as surface tension. Further, it is directed opposite surface tension when rising above the water level, thus cooperating with the wing lift to help the insect rise higher above the interface and break the meniscus. This reduces the wing lift required to counter surface tension and body weight and take off from the interface into air. Since the oscillations responsible for assisting take-off arise from the balance between wing force and the non-linear resultant of surface tension, it is expected that the relationship between take-off times and the wing forcing parameters \textendash\space wingbeat frequency $f$, initial stroke phase angle $\phi$, and ratio of force produced in upstroke and downstroke $r$ \textendash\space is complex and non-monotonic. Using a computational survey of parameter space, we explored the effect of these wingbeat parameters on the takeoff time for an insect in interfacial flight, over a range of lift force magnitudes. This virtual parameter sweep allows us to study variations in individual flight parameters that it is not possible to independently or physiologically manipulate in the real world. We have explored the effects of changing the wingbeat frequency over a wide range both above and below the typical frequency at room temperature (Fig. 6C), the effects of changing the ratios of upstroke force to downstroke force (Fig. 6D) and even the effects of making the legs thicker or thinner (Fig. S4). This is clearly extremely hard to achieve in the real world, with a single species of insect or even to make a fair comparison between different species based on single-parameter variations, all other factors kept constant. Hence, we leverage the power of mathematics and numerical simulations in this section with the aim of making single-parameter sweeps where all other factors are retained constant, in situations that are not physiologically accessible easily. (Takeoff times for 2-leg flight are shown in Fig. 6C,D,E. Takeoff times for 4-leg flight are shown in Fig. S2A,B,C.) Broadly, we see that isochronous contours for take-off time have very complex shapes, and that take-off is possible even below the static lift requirements. In the absence of effects like body angle variation and jumping, while the insect can take off from the interface instantly at high wing lifts, it remains trapped on the interface at lower lift magnitudes. In the intermediate regimes of lift $L\approx L_{max}$ where \textit{G.nymphaeae} operates, it takes a certain amount of time to build up energy in meniscus oscillations and takes off from the interface with some delay. While take-off times range from a few milliseconds to a few seconds, we note that a significant proportion of trajectories have take-off times of the order of several wingbeats (tens of milliseconds). This is of the same order of time as the neural responses involved in active flight control in insects \cite{ristrophwangcohen}. For trajectories in these regimes, we infer that there is no active control of flight during take-off. It is therefore desirable that for passive stabilization, perturbations of the interfacial flight trajectory are not amplified. For the trajectory to remain stable during take-off, it is preferable for the system dynamics to be robust to disturbances.\\

\noindent\textbf{Surface tension drives chaotic oscillatory trajectories normal to the interface}\\ 
We analyzed the stability of interfacial flight dynamics to see whether small perturbations in the trajectory are damped out or amplified to a large extent. We rewrote the forced non-linear oscillator model for the vertical component of interfacial flight as a set of three first-order ODEs for the variables $\{t,y,\dot{y}\}$. On linearizing the system of equations for small perturbations from rest, we found that the solutions have Lyapunov exponents with signs $\{0,+,-\}$. The Lyapunov exponents are an indicator of the rate of change of a given trajectory in phase space along the dimensions represented by the chosen variables $\{t,y,\dot{y}\}$, and the signs obtained imply divergence of perturbations along one dimension with a contraction in the other, which is characteristic of chaotic phase space trajectories \cite{dingwell}. A rigorous analytical derivation of Lyapunov exponents is provided in the Supplementary Information. To study the full nonlinear behaviour, we constructed a bifurcation diagram to see how the periodicity of interfacial flight trajectories varies with the wing lift amplitude that forces the vertical oscillations (Fig. 7A). For an insect where both hindlegs and forelegs remain in contact with the interface throughout, there is initially a periodic regime at low values of the lift-to-weight ratio $q$. As $q$ increases, several bifurcations lead to multi-periodic orbits, with aperiodic regimes beginning to appear between islands of periodic orbits as the $q$ approaches $1$ \textemdash\space the minimum value of $q$ needed to support the insect's body weight in air. These orbits include small period-three regions between $q=1$ and $1.1$, which are a definite indicator of chaos in the system. There is a wide chaotic belt between $q\approx1.1$ and $2.8$, where most modern insects capable of airborne flight would be found. Trajectories begin to quickly take off from the interface as lift amplitudes exceed $q\approx2.8$. We looked at phase plots for two trajectories from the periodic and chaotic regimes, which show complex shapes when embedded in the $\{y, \dot{y}\}$ plane (Fig. 7B). It is interesting to note that the structure of the phase plot from the chaotic interfacial flight regime with $q=1.48$ (Fig. 7B lower panel) strongly resembles the strange attractor in another chaotic surface tension based oscillator \textemdash\space the soap-film fluid trampoline \cite{giletbush} \textemdash\space even though the physical mechanisms and mathematical models for chaotic oscillations are different for the two systems. Trajectories in the chaotic regime also display a sensitive dependence on initial conditions as shown in Fig. 7C, which is the hallmark of chaos in the system. To confirm that our mathematical prediction of chaos applies to experimentally recorded trajectories, we constructed a delay plot (Fig. 7D) for the representative trajectory shown in Fig. 3A. Fig. 7D shows that the vertical displacement in this trajectory is completely uncorrelated across successive wingbeats, without any periodic or repeating structures. Since the errors due to measurement are much smaller than the length scales for displacement, we conclude that the lack of correlation is truly a consequence of chaos and not an artefact of experimental noise. These results indicate that small fluctuations in an interfacial flight trajectory could lead to significant instabilities, hence it is not possible to predict when the insect takes off from the interface in a chaotic regime. Indeed, we observe no correlations in our experiments between the take-off times and kinematic parameters of the flight trajectory such as velocity (data not shown).\\\\

\noindent\textbf{DISCUSSION}\\
This work provides the first biomechanical analysis of interfacial flight in any insect. The waterlily beetle \textit{G.nymphaeae} we have studied is also capable of well-developed airborne flight, which points towards the fact that interfacial flight in these beetles is a derived adaptation to their aquatic ecological niche on the surface of a pond. We have demonstrated the dominant influence and unique role of interfacial forces like capillary gravity wave drag and surface tension in interfacial flight. These nonlinear forces add a hitherto unforeseen complexity to interfacial flight and differentiate it from conventional airborne flight.\\

Using high speed videos of \textit{G.nymphaeae} flight kinematics, we have developed a quantitative model to capture the fundamental physics underpinning interfacial flight. Our analysis shows that at typical speeds, interfacial flight actually requires stronger wing thrust than airborne flight at the same speed, due to the high capillary gravity wave drag. We however note that \textit{G.nymphaeae} clearly shows a preference for fast interfacial flight over airborne flight despite the higher energy expenditure. We suggest that the searching efficiency provided by two-dimensional locomotion in foraging for floating food on a planar interface is sufficiently high to give this beetle a distinct advantage in staying on the interface. We speculate that another advantage of fast locomotion is to avoid underwater predators like fish, that have been known to detect capillary waves on the surface of a pond generated by trapped insects that fall on the fluid surface \cite{schwarz}. Another fascinating observation in many experiments is the lack of any active braking mechanism in 2D flight (Movie SV11). We propose that the insect actually takes advantage of the high drag forces on the interface to passively maneuver, slow down or stop its motion during interfacial flight. Next, we have shown that maneuvers in the transition between different flight modes - interfacial, airborne and backward, are controlled by postural changes. The body angle is the most probable tuning parameter used by the insect, being strongly correlated to the stroke plane angle and altering the ratio of lift and thrust produced by the wings.\\

Third, using both successful (Movie SV12) and failed (Movie SV9) takeoff attempts, we demonstrated that surface tension acts as an energy trap that necessitates enhanced lift production to take off from the interface. Takeoff from terrestrial surface into air is itself a complex and intriguing phenomenon in insect flight, needing lift-producing mechanisms beyond wing flapping. Previous studies on insect takeoff have shown roles for ground effect and leg extension in producing lift \cite{bimbard}. The takeoff of \textit{G.nymphaeae} from a fluid interface into air gains further complexity, as there are additional surface tension forces tethering the insect to the interface. These surface tension forces are large, being several times the body weight at the maximum extent. This argues against the hypothesis of a transition between interfacial and airborne flight simply by increasing the wing forces produced by insects. However, it highlights the importance of additional take-off mechanisms used by insects as observed by Marden et al. and Bimbard et al.,  such as reducing the number of legs in contact with water or jumping off the surface. An important possible contribution to further enhancing the wing lift produced by an insect in interfacial flight is the ground effect \cite{rayner}. The insect beats its wings barely a few millimetres above the surface of water. The reflection of wingtip vortices from the water surface can increase the unsteady lift produced by the wings, for the same biomechanical energy cost. In future work, we intend to investigate this novel vortex induced ground effect and its role in lift generation. \textit{G.nymphaeae} is an ideal model organism for future studies in this area.\\

Fourth, we have shown that interfacial fliers are assisted in breaking the meniscus and taking off from the interface by dynamic energy storage in vertical oscillations of the meniscus. The optimal conditions for minimizing either the time or the lift required for take-off are different, and have an intricate dependence on the wingbeat parameters. It is interesting to note that typical values of some of these parameters are far from optimal in \textit{G.nymphaeae}. For example, the optimal frequency to take off from the interface using the minimum wing lift is $60$ Hz for an insect in 4-leg interfacial flight (Fig. 6C), and about $30$ Hz for an insect in 2-leg flight (Fig. S2A). However, the physiological value of wingbeat frequency is constant at about $116$ Hz (Fig. S1A), which is far removed from these optimal frequencies. These oscillations of the meniscus can often lead to take-off at timescales that are comparable to the time required to mount an active neural response for flight control. This can pose a challenge in controlling the take-off process, particularly when disturbances in the trajectory are amplified.\\

Finally, we have shown that interfacial flight is an example of chaos arising naturally in a biological system. The dependence of biological functions on chaotic rhythms or patterns has been demonstrated in systems as diverse as vortex interactions with flapping wings \cite{lentink}, tidal bobbing in giant kelp \cite{denny}, amputee runners adjusting their stability \cite{look}, circadian rhythms in foraging ants \cite{nicolis}, TNF-driven inflammatory responses in cells \cite{jensenkrishna}, and turbulent flow driven maturation in sea urchin larvae \cite{gaylord}. In the context of interfacial flight dynamics, the emergence of chaotic trajectories hints at the challenge of developing robust flight control mechanisms on the interface, since higher lift magnitudes produce inherently unstable trajectories. This suggests that evolving finer flight control mechanisms in tandem with stronger wings could be difficult on an air-water interface. Our work opens up interesting possibilities for designing new kinds of robust control algorithms for bio-mimetic robots that operate on fluid interfaces.\\

Our discovery of complex interfacial flight behaviour in \textit{G.nymphaeae} provides an excellent experimental platform to study the ground effect in insect flight, the kinematics and dynamics of the progression from interfacial to airborne flight, neural responses in a chaotic biological system, and new kinds of interfacial oscillators. The model of interfacial flight that we have presented provides a starting point for quantitatively understanding the basic physical phenomena involved. This can be used as the basis for further refinement and also future studies on varied forms of interfacial flight and surface skimming in different organisms and robotic systems, with more detailed and organism-specific expressions for the individual forces involved. We hope that our work will open up new avenues for both experimental and analytical investigations into insect behaviour, biomechanics, robotics and interfacial fluid mechanics.\\\\

\noindent\textbf{Competing Interests: }  The authors declare that they have no competing financial interests.\\\\
\noindent\textbf{Author Contributions: } M.P. discovered 2D flight in beetles in the field. T.B., D.-H.K., H.M. and M.P collected further field and laboratory data. H.M. and M.P analyzed the data, developed the dynamical system model and wrote the manuscript.\\\\
\noindent\textbf{Acknowledgements: } We acknowledge all members of the Prakash Lab at Stanford University for meaningful advice and insights. We specifically acknowledge high school student Aditya Gande, who wrote the code for automated tracking of insect flight. We also acknowledge Harvard Forest for providing a place to collect insects, and Harvard Society of Fellows for providing a place to freely contemplate some of the ideas presented here. Finally, we acknowledge our reviewers for their insightful comments on our manuscript.\\\\
\noindent\textbf{Funding: } H.M. acknowledges support from the Howard Hughes Medical Institute International Student Research Fellowship. M.P acknowledges funding from the NSF Career Award and the Pew Foundation Fellowship. 

\newpage
\noindent\textbf{References : }
\begingroup
\renewcommand{\section}[2]{}

\newpage
\noindent\textbf{FIGURE 1: Interfacial flight in \textit{Galerucella nymphaeae}.}\\
\noindent\textbf{A}, Natural habitat of \textit{G.nymphaeae} in Harvard Forest, MA, where the first specimens were observed and captured. Inset shows a wild specimen resting on a lily leaf. \textbf{B}, Close-up of \textit{G.nymphaeae}. Inset shows the leg resting on a drop of water, such that the tarsi are unwetted and supported on the drop while the claws at the tip are immersed in the water. Note that the drop is deformed near the claws. \textbf{C}, Capillary waves are generated during 2D flight, due to the perturbation of the interface by the insect's immersed claws. \textbf{D}, Schematic illustration of interfacial flight in \textit{G.nymphaeae}. \textbf{E} (side view), \textbf{F} (rear view), \textit{G.nymphaeae} posture in upstroke, midtroke and downstroke of interfacial flight. The middle pair of legs is raised above the body and the body is angled such that its weight is well supported between the four immersed legs. White scale bar is $1$ mm. Black scale bars are $5$ mm.\\

\noindent\textbf{FIGURE 2. Adaptations enabling interfacial flight in \textit{G.nymphaeae}}\\
\noindent\textbf{A}, False colour SEM image indicating wetting and non-wetting regions on the leg. Green indicates superhydrophobic regions and blue indicates hydrophilic regions. \textbf{B}, SEM images of \textit{G.nymphaeae} body and legs showing hydrophobic hairy structures. \textbf{C}, \textbf{D}, \textbf{E}, Successive magnifications of the hind leg ultrastructure \textbf{(D)} showing tarsi with hydrophobic setae \textbf{(E)}, and a pair of curved, hydrophilic claws \textbf{(C)} which are immersed. \textbf{F-H}, Similar hydrophobic-hydrophilic ultrastructural line barriers seen on hind legs \textbf{(F)} and forelegs \textbf{(G, H)} of more beetles. \textbf{J}, Schematic showing pinning of contact line at tarsal claws during interfacial flight. \textbf{K}, Meniscus formed by dipping leg of dead beetle into water and raising up to the maximum extent. The white curve is the computed theoretical minimal surface profile, which fits the experimentally produced meniscus well. \textbf{L}, Sequence showing formation and breakage of meniscus at the claw during takeoff in a live beetle. White scale bars are $100\thinspace\mu$m. Black scale bar is $5$ mm.\\

\noindent\textbf{FIGURE 3. Kinematics of interfacial flight trajectories}\\
\noindent\textbf{A}, Representative interfacial flight trajectory showing vertical oscillations. Inset shows a snapshot from the video used to generate the trajectory, with the arrow pointing to the femur-tibia joint of the insect's hind leg - a natural marker used for tracking. Error bars were calculated for tracked coordinates as the pixel resolution of the video, equal to approximately $10\thinspace\mu$m per pixel. \textbf{B}, Horizontal displacement increases rapidly, with average displacement per wingbeat varying quadratically with time. Inset shows the displacement during a single wingbeat, with steep increase during downstroke and sigmoidal variation during upstroke. \textbf{C}, Variation in horizontal velocity over time showing a linear increase in the average downstroke velocity. Velocity is maximum in the downstroke part of a wingbeat and drops to a minimum in mid-upstroke before recovering. Velocities were computed by fitting a spline to the displacement and calculating its slope. Error bars were derived from displacement errors.\\

\noindent\textbf{FIGURE 4. Dynamic model for interfacial flight}\\
\noindent\textbf{A}, Schematic depicting the different forces acting on the beetle. Boxed insets on the right show the beetle reduced to a single particle pinned at the interface, with horizontal forces (top) and vertical forces (bottom) acting on the particle. Circled insects on the left show the direction reversal of surface tension, depending on the nature of deformation of the meniscus at the pinned contact line.\\

\noindent\textbf{FIGURE 5. Experimentally derived dynamics of wing forces exerted during interfacial flight}\\
\noindent \textbf{A}, Stroke plane decreases linearly as body angle increases during flight. Increasing body angle leads to transitions between interfacial, airborne or backward flight modes. Error bars are estimated from pixel resolution error in extreme wing and body points used to calculate angles. \textbf{B}, As an insect takes off from the interface into air, body angle steadily increases while stroke plane angle approaches zero. When stroke plane angle falls below zero, the flight direction is backwards, opposite the dorsal-to-ventral axis. \textbf{C-E}, Failed takeoff attempt where legs are trapped by surface tension. In \textbf{C}, a sequence of images shows a failed take-off attempt, where a significant portion of the legs is wetted. Note the almost vertical posture (body angle $\sim 90^{\circ}$) that concentrates wing force into lift. Black scale bar is 5mm. \textbf{D}, Horizontal displacement and body angle both increase initially, and then level off as the thrust approaches zero (yellow shaded region on the right.) Error bars indicate pixel resolution. \textbf{E}, Mean horizontal velocity in each wingbeat is almost zero, resulting in a static equilibrium. Error bars were derived from positional errors corresponding to one pixel.\\

\noindent\textbf{FIGURE 6. Computational modelling of capillary drag, oscillations and takeoff}\\
\noindent\textbf{A}, Comparison showing that total drag experienced in interfacial flight is much larger than that in airborne flight, above a critical velocity. The additional capillary-gravity wave drag experienced in interfacial flight is shown as the pink shaded region. \textbf{B}, Different trajectories computed for a given total wing force magnitude, by varying the wing force distribution between lift and thrust. Trajectories take off for lower thrust-to-lift ratio $p$ and remain confined to the interface for higher $p$. Dashed line is the mean water level at infinity and dotted line is the maximum vertical height of meniscus attachment ($260\thinspace\mu$m). \textbf{C-E}, Phase plots for variation in take-off time for different values of lift force per unit body weight, for an insect with 2 legs immersed in water. The single other parameter varied in each plot is wingbeat frequency in (\textbf{C}), symmetry ratio ($r$) between wing force produced in upstroke and downstroke in (\textbf{D}), and wing stroke angle at initiation of motion in (\textbf{E}). White dotted lines show maximum estimated lift $L_{max}=2*(2\pi R_{claw}\sigma)+mg$ and wingbeat frequency $f_{w}\approx 116$ Hz.\\

\noindent\textbf{FIGURE 7. Chaotic vertical oscillations on the interface} \\
\noindent\textbf{A}, Bifurcation diagram for trajectories at different lift-to-weight ratio $q$. Top - A wide range of $q$ shows $5$ distinct trajectory regimes – (i) Periodic and confined to interface, period equal to or half the wingbeat period (ii) Chaotic, confined to the interface (iii) Chaotic, takes off from the interface with some time delay (iv) Periodic and confined to interface, period equal to wingbeat period (v) Instantaneous take-off. Bottom - Close-up of transition between regimes (i) and (ii) showing cascades and regions of chaos. \textbf{B}, Phase plot at $q=0.99$ (top) shows a self-intersecting period-4 cycle. Phase plot at $q=1.48$ (bottom) shows a chaotic attractor also seen in other oscillators driven by surface tension \cite{giletbush}. \textbf{C}, Divergence of two trajectories with closely spaced initial conditions in the chaotic regime ($q=1.48$), where one takes off and the other is trapped on the interface. Both trajectories are simulated with parameters $L=137\mu N$, $f=116 Hz$, $r=0.15$, $\phi =\pi/2$, but for different initial conditions which do not lie on each other's $\{y,\dot{y}\}$ phase plot. \textbf{D}, Delay plot of experimental trajectory data showing vertical displacement plotted against itself with a delay of $\tau=1$ wingbeat $\approx8.67$ ms. The lack of any repeated structure indicates that the vertical oscillations observed in the experimentally recorded trajectory do not have any correlation in time and are chaotic in nature.

\begin{figure}[H]
        \center{\includegraphics[width=\textwidth]
        {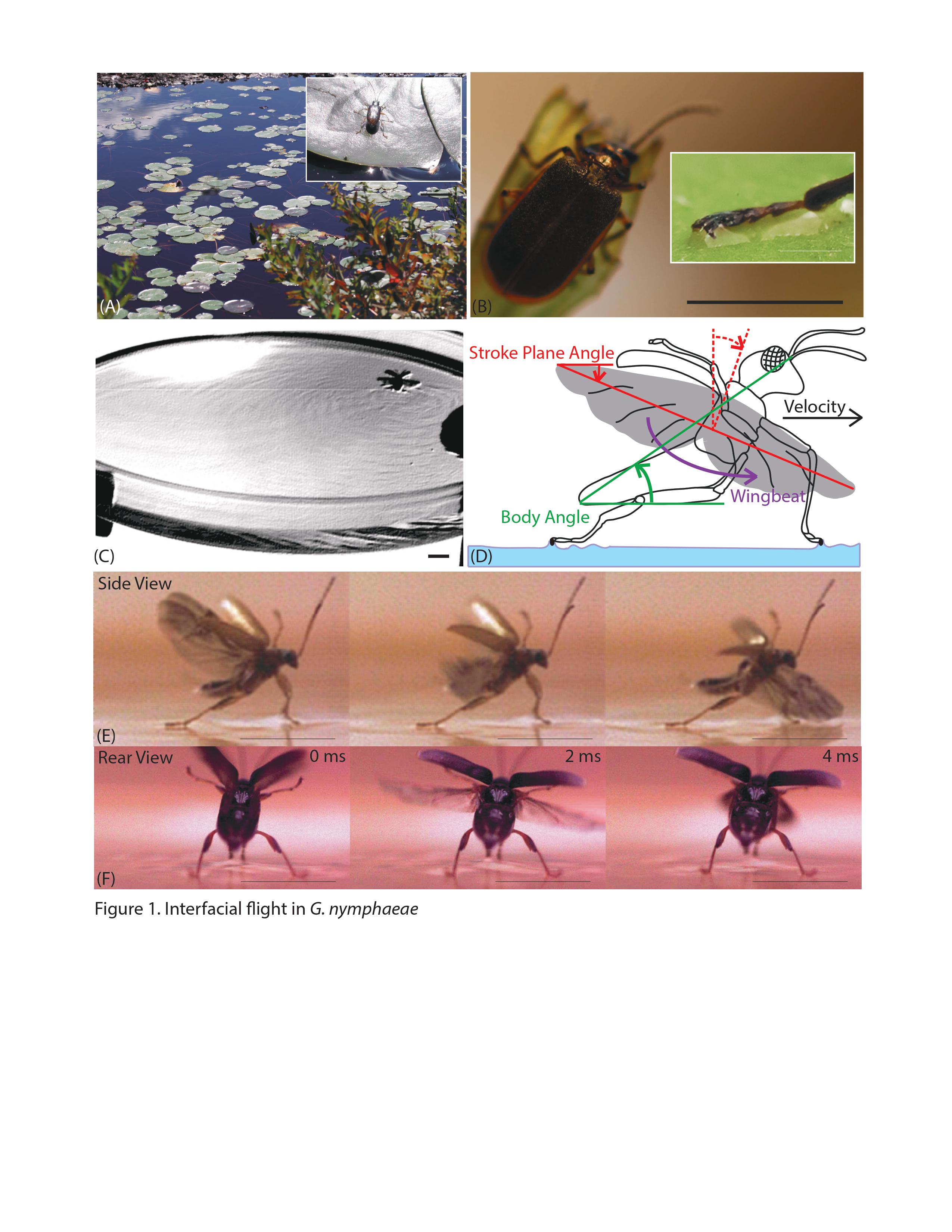}}
\end{figure}

\begin{figure}[H]
        \center{\includegraphics[width=\textwidth]
        {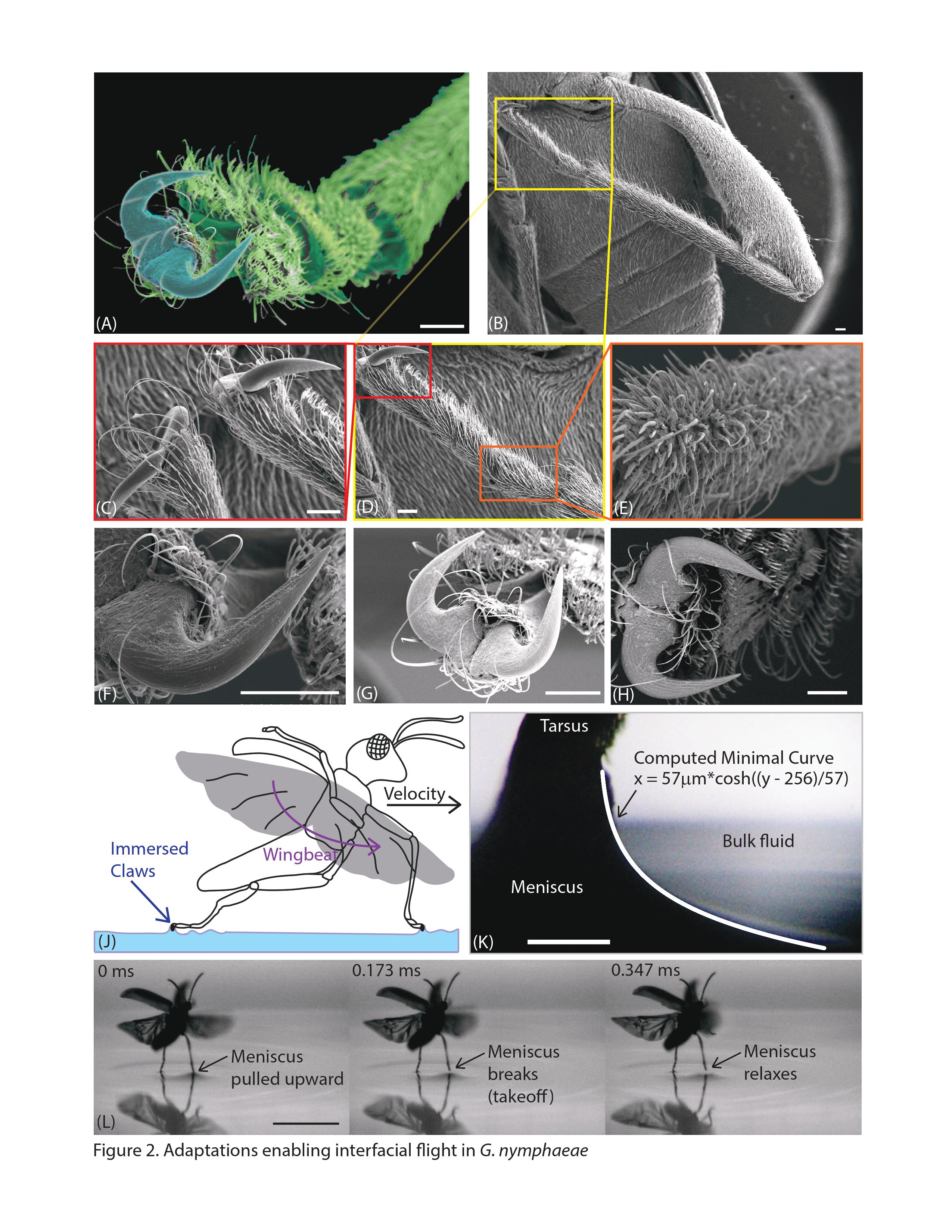}}
\end{figure}

\begin{figure}[H]
        \center{\includegraphics[width=\textwidth]
        {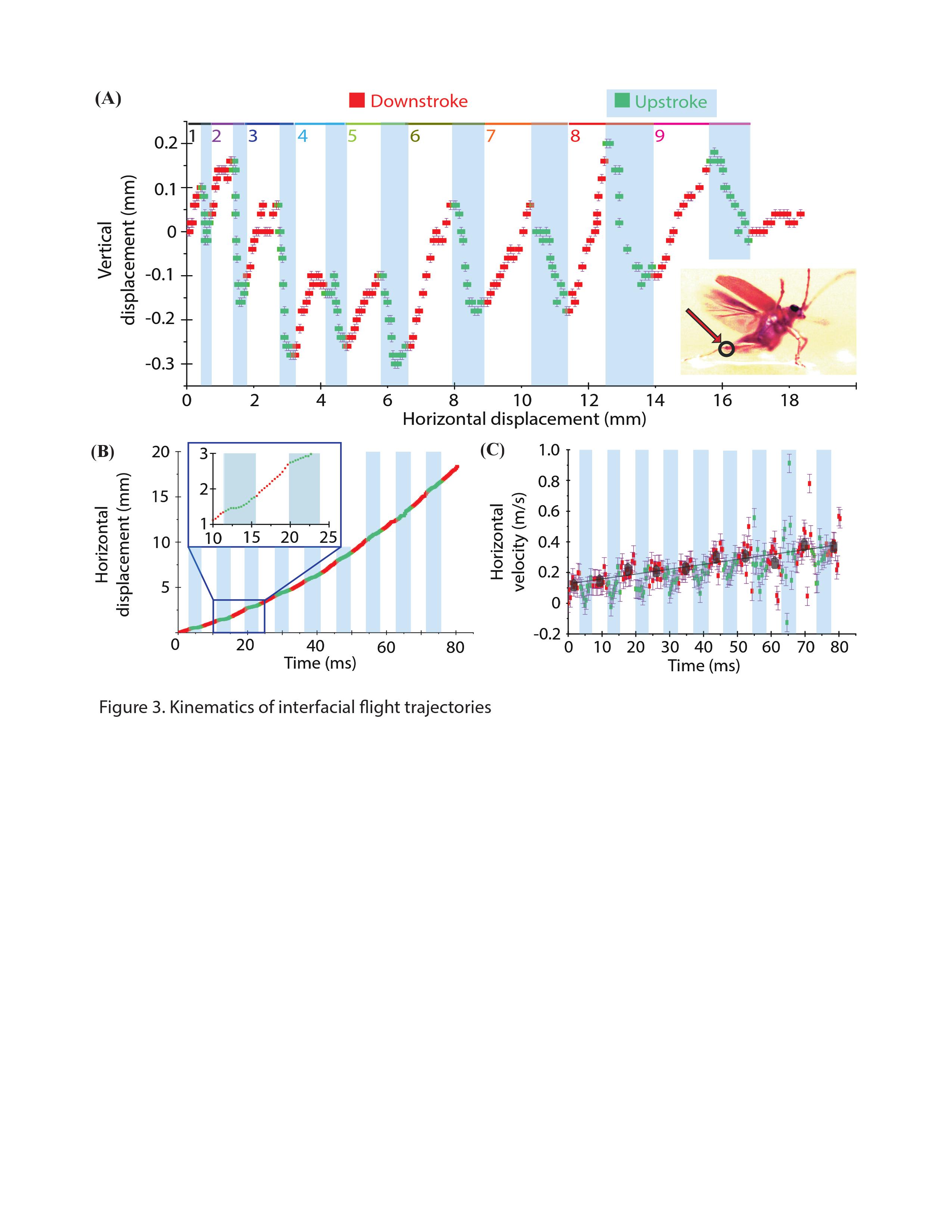}}
\end{figure}

\begin{figure}[H]
        \center{\includegraphics[width=\textwidth]
        {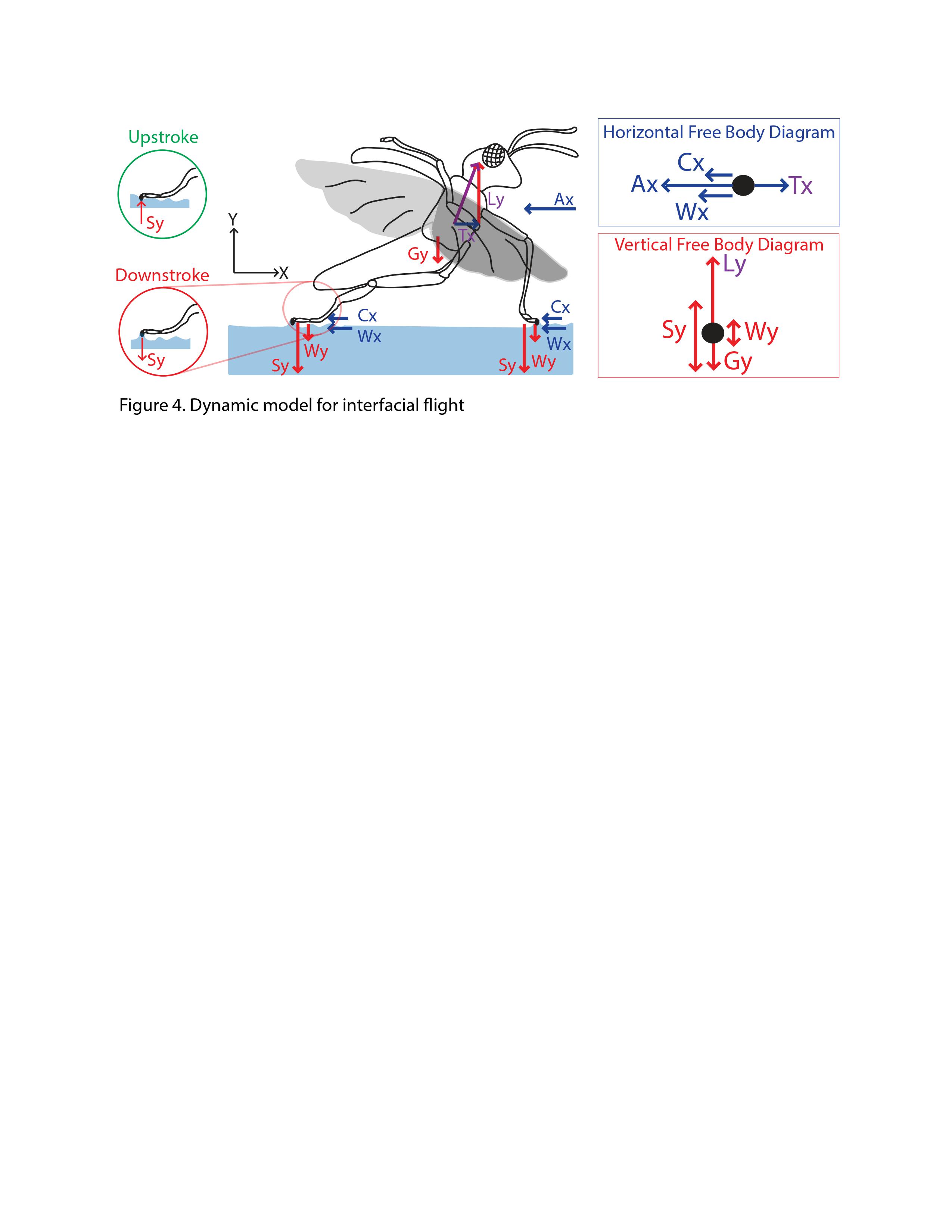}}
\end{figure}

\begin{figure}[H]
        \center{\includegraphics[width=\textwidth]
        {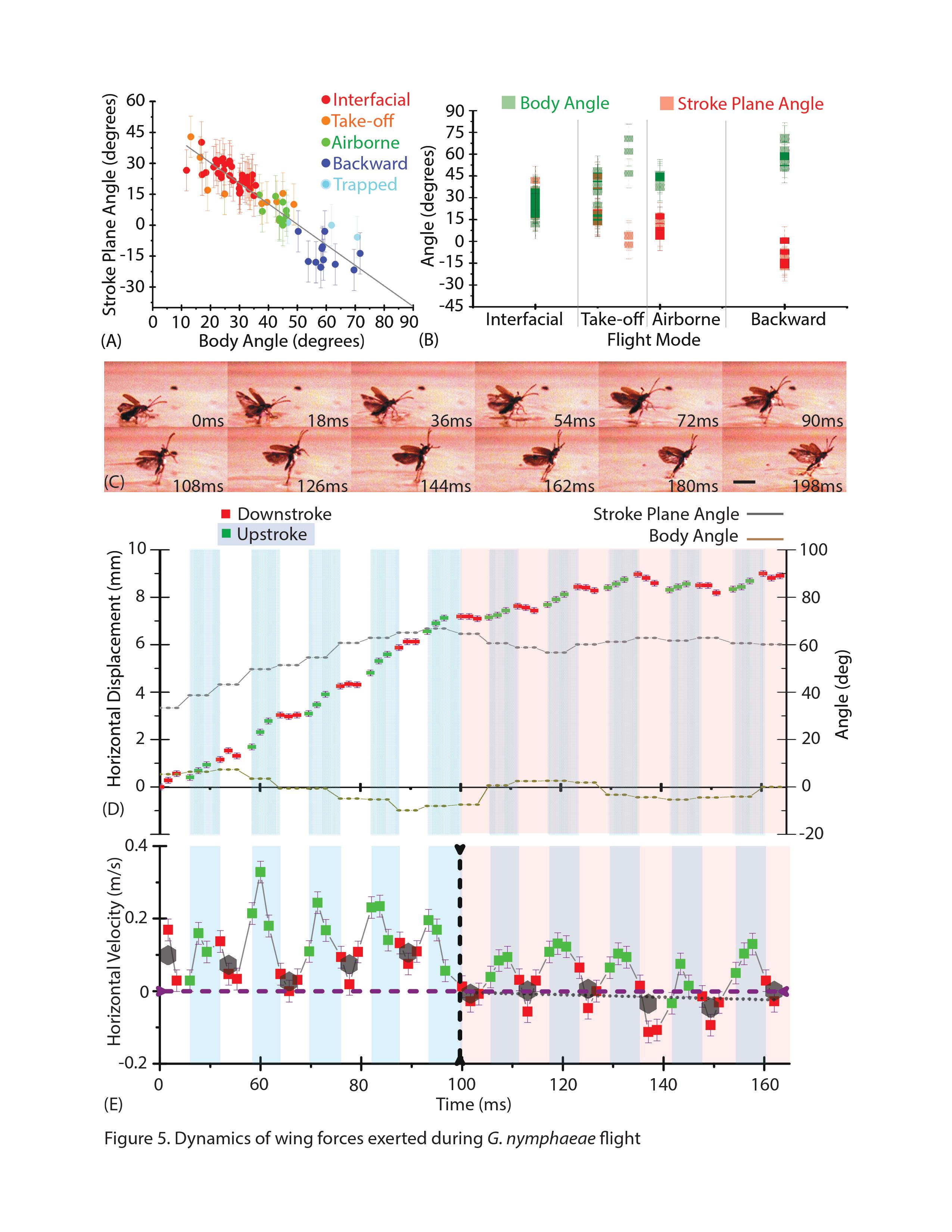}}
\end{figure}

\begin{figure}[H]
        \center{\includegraphics[width=\textwidth]
        {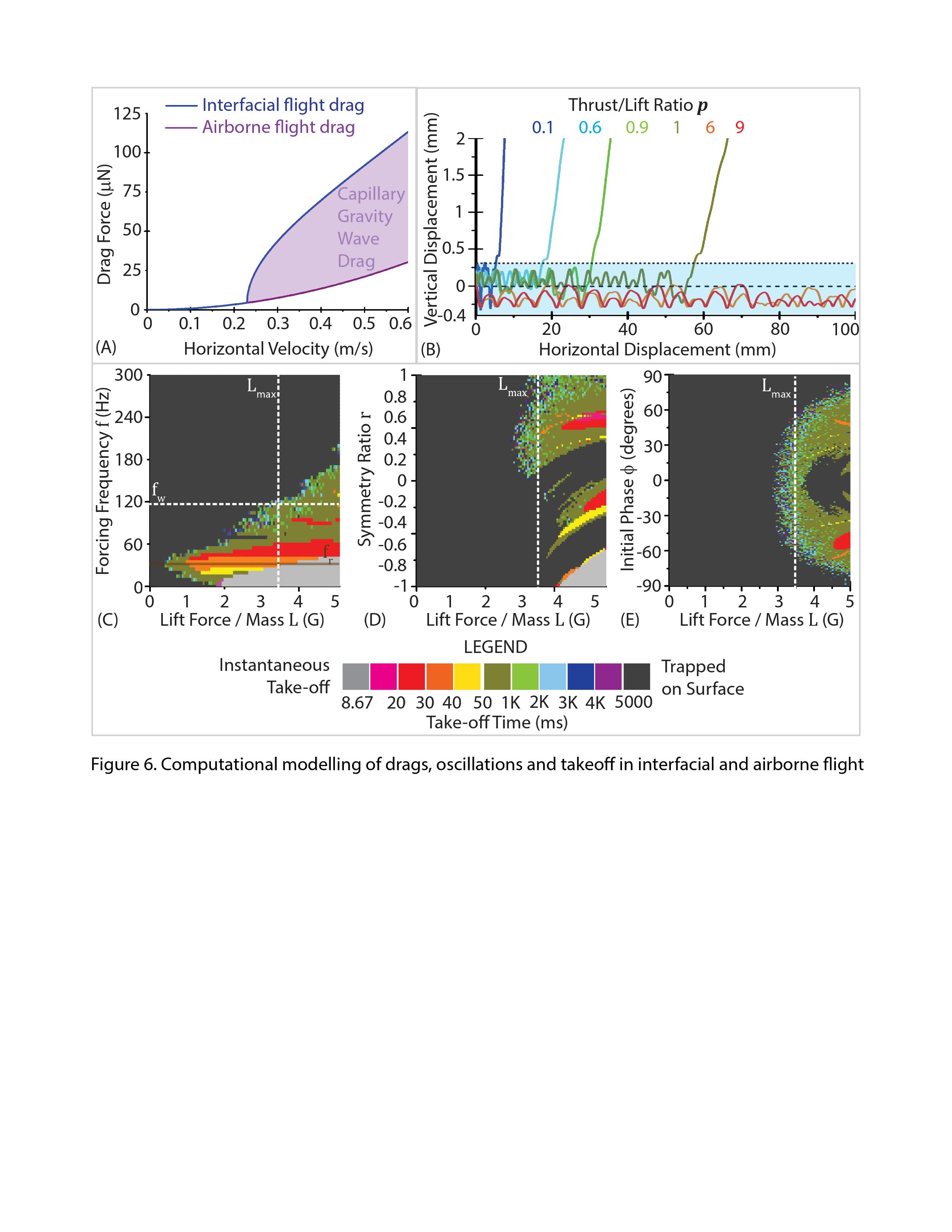}}
\end{figure}

\begin{figure}[H]
        \center{\includegraphics[width=\textwidth]
        {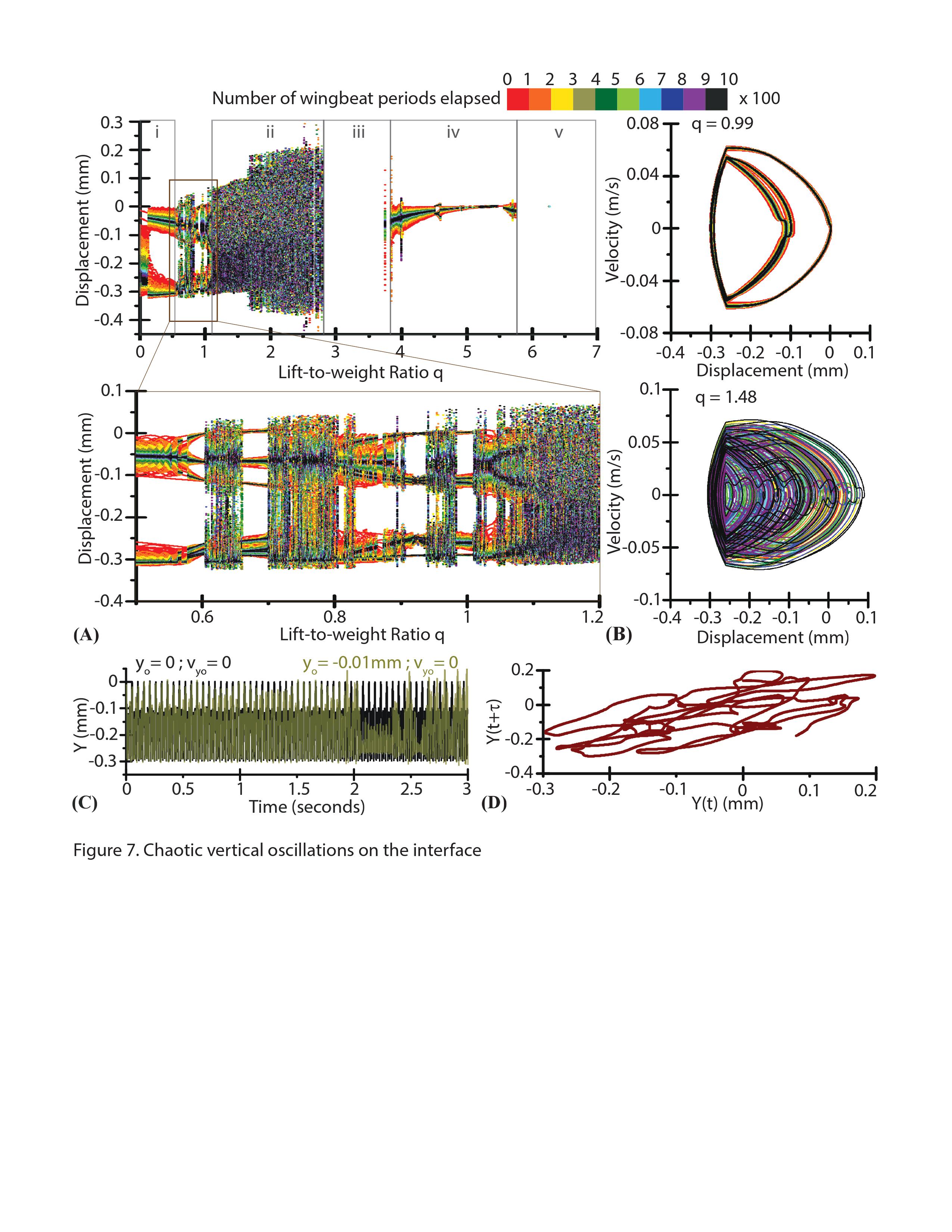}}
\end{figure}

\newpage
\noindent\textbf{SUPPLEMENTARY INFORMATION}\\\\
\noindent\textbf{Supplementary figures S1 to S4}\\\\

\noindent\textbf{SUPPLEMENTARY FIGURE 1. Kinematics of additional interfacial flight trajectories}\\
\noindent\textbf{A}, Plot of wingbeat period across wingbeats. Error bars are derived assuming deviations of one frame duration. \textbf{B}, Plot of wing joint angle across flight modes, calculated as sum of stroke plane angle and body angle, showing low variation across different flight modes. Error bars are estimated from pixel resolution error in extreme wing and body points used to calculate angles. \textbf{C,D}, Trajectories for additional flight sequences showing both interfacial flight with oscillations and airborne flight. Errors bars correspond to pixel resolution. \textbf{E}, Plot of trajectory for flight sequence as extracted at head, mouthparts and hind leg. \textbf{F}, Close-up showing pronounced discrepancies between trajectories measured at head and hind leg for interfacial flight. \textbf{G}, Average horizontal velocity for a flight trajectory, showing different accelerations (slopes) below and above the threshold velocity $c_{min}=0.23$ ms$^{-1}$ for the onset of capillary gravity wave drag.\\
\begin{figure}[H] 
        \center{\includegraphics[width=\textwidth]
        {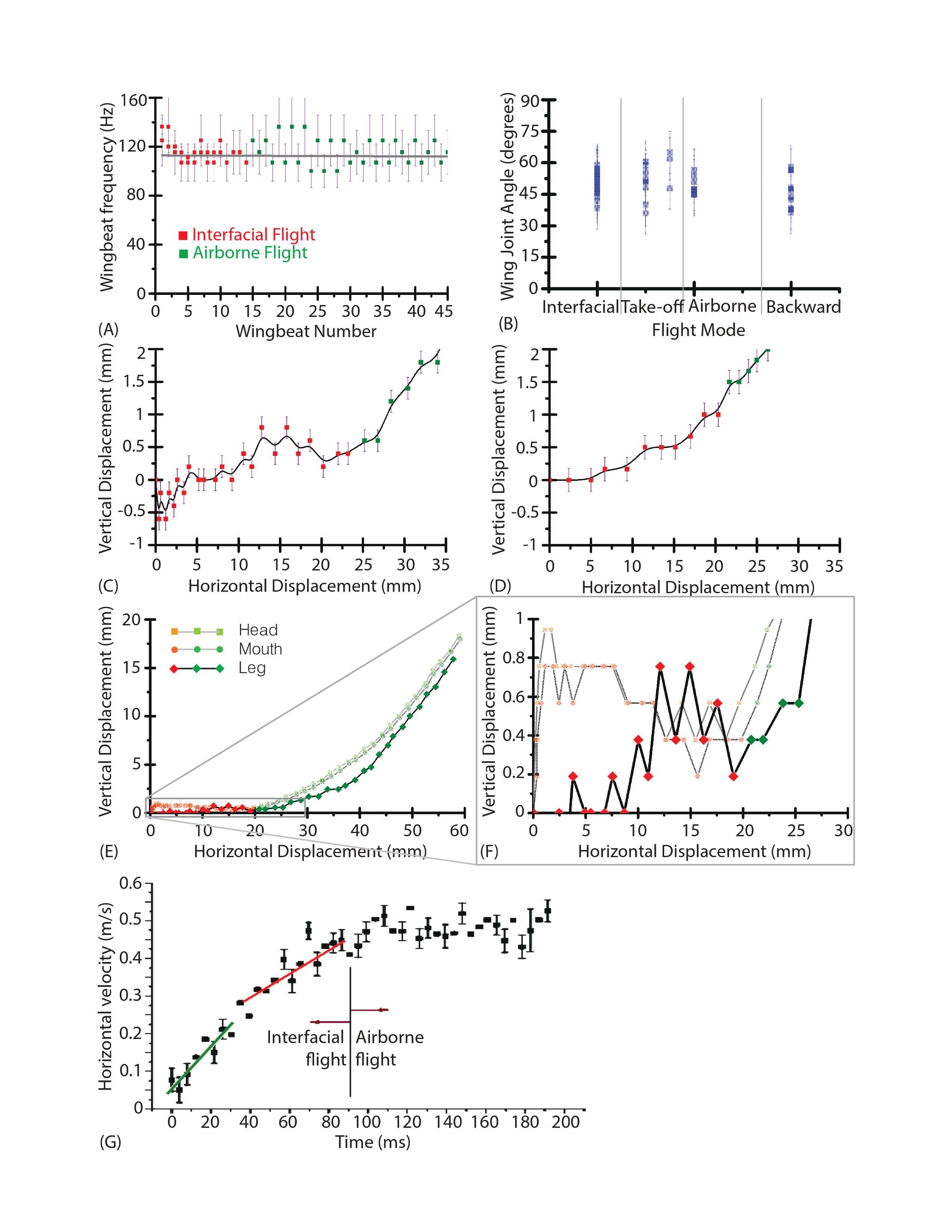}}
\end{figure}

\newpage
\noindent\textbf{SUPPLEMENTARY FIGURE 2. Variation of take-off time with wing forcing parameters for 4-leg flight}\\
\textbf{A-C}, Phase plots for variation in take-off time for different values of lift force per unit body weight. The single other parameter varied in each plot is wingbeat frequency in (\textbf{A}), symmetry ratio ($r$) between wing force produced in upstroke and downstroke in (\textbf{B}), and wing stroke angle at initiation of motion in (\textbf{C}). White dotted lines show maximum estimated lift $L_{\sigma}+1=2*(2\pi R_{claw}\sigma)+mg$ and wingbeat frequency $f_{w}$.\\\\
\begin{figure}[H] 
        \center{\includegraphics[width=0.9\textwidth]
        {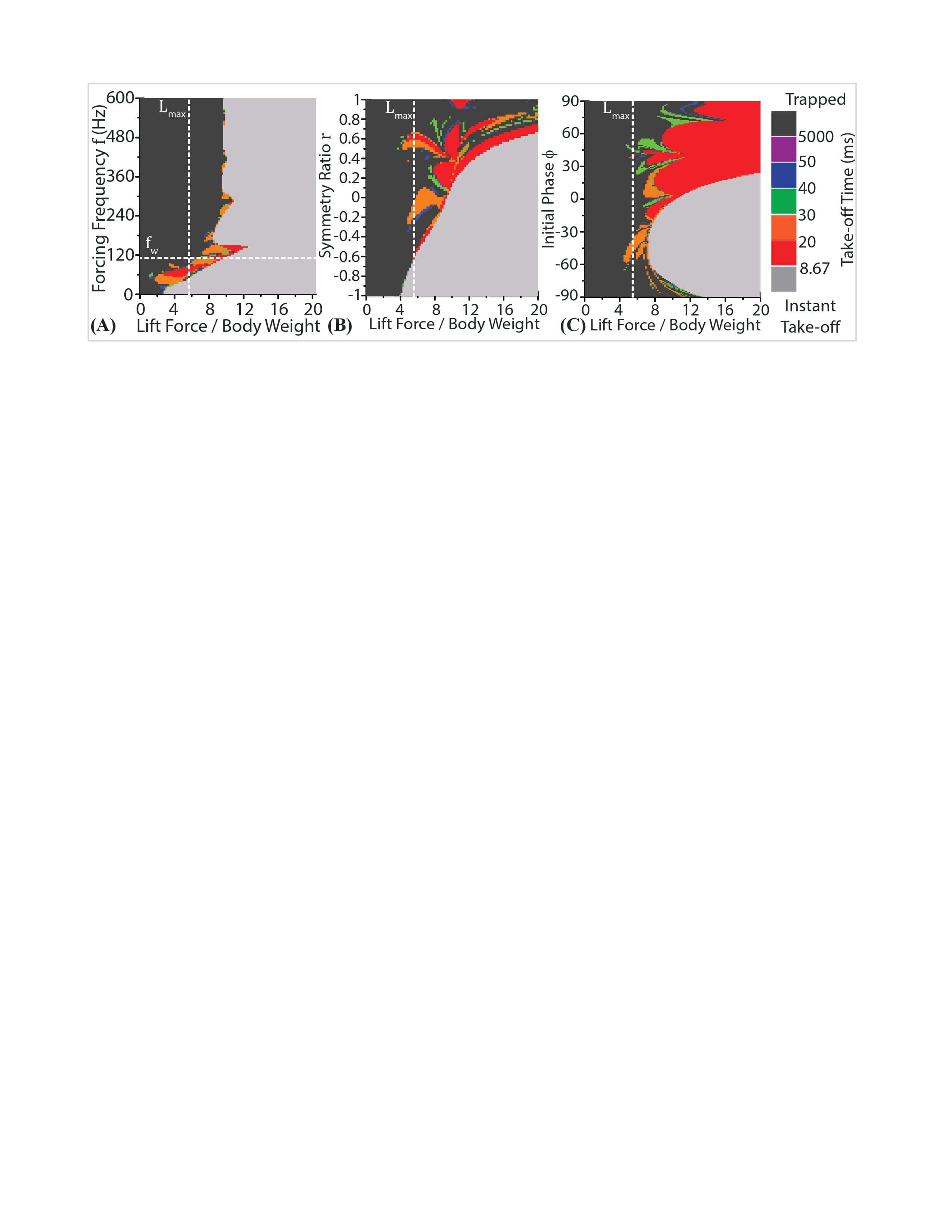}}
\end{figure}

\newpage
\noindent\textbf{SUPPLEMENTARY FIGURE 3. Validation of dynamic model by matching major kinematics trends}\\
Simulated trends are shown in violet and superimposed on experimental data shown in Figure 3. \textbf{A}, Trends are matched in horizontal displacement covered in $9$ wingbeats, and the presence of vertical oscillations of varying peak level and amplitude of the same extent as meniscus deformation. \textbf{B}, Horizontal displacement in simulated trajectory closely matches experimentally obtained data. \textbf{C}, Horizontal velocity in simulated trajectory has similar magnitude and trends as experimentally obtained data.\\
\begin{figure}[H] 
        \center{\includegraphics[width=0.9\textwidth]
        {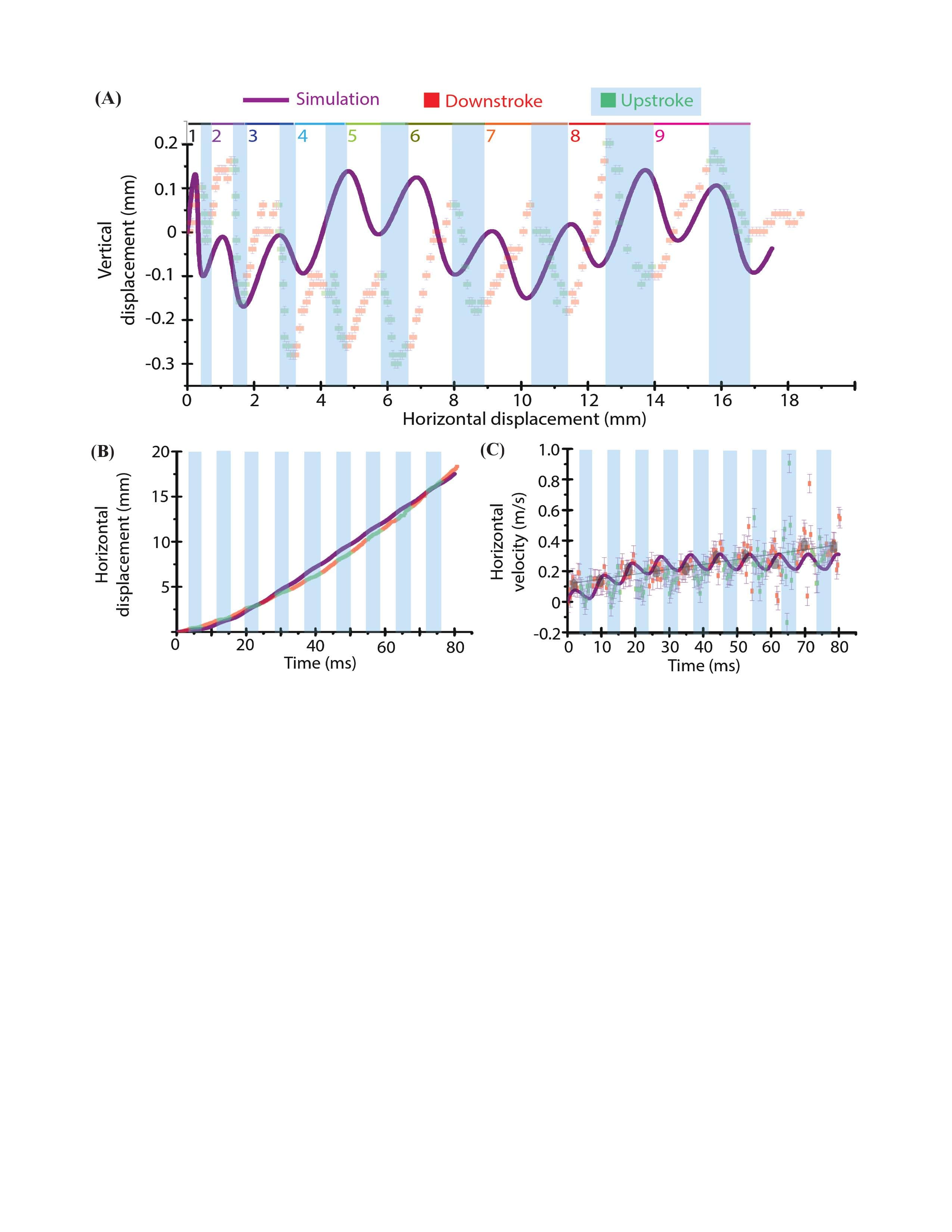}}
\end{figure}

\newpage
\noindent\textbf{SUPPLEMENTARY FIGURE 4. Effect of varying thrust-to-lift ratio and leg radius (adhesion due to surface tension)}\\
\textbf{A}, Computational modeling of trajectory variations due to \textbf{A}, different thrust-to-lift ratio at constant surface tension adhesion (leg radius) and \textbf{B}, varying surface tension adhesion (leg radius) at constant lift and thrust. It can be seen that when surface tension is reduced, the insect needs far less lift to take off from the surface.  However, it also does not cover much horizontal distance as it simply doesn’t stay on the surface long enough. It can remain on the surface by reducing the total wing force, but this would not enable it to develop high velocities along the interface. A strategy which would allow the insect to skim quickly for longer distance along the surfaces and then take off is by channeling the entire wing force into thrust with very little lift, which gives rise to horizontal motion. Changing the wing angle to increase lift gives rise to takeoff, as indicated by Fig. 5A where flight mode is strongly correlated with stroke plane angle. It is important to note further that such a mechanism is independent of the actual values of lift and adhesion from surface tension, depending simply on the relative magnitudes of the two.\\
\begin{figure}[H] 
        \center{\includegraphics[width=\textwidth]
        {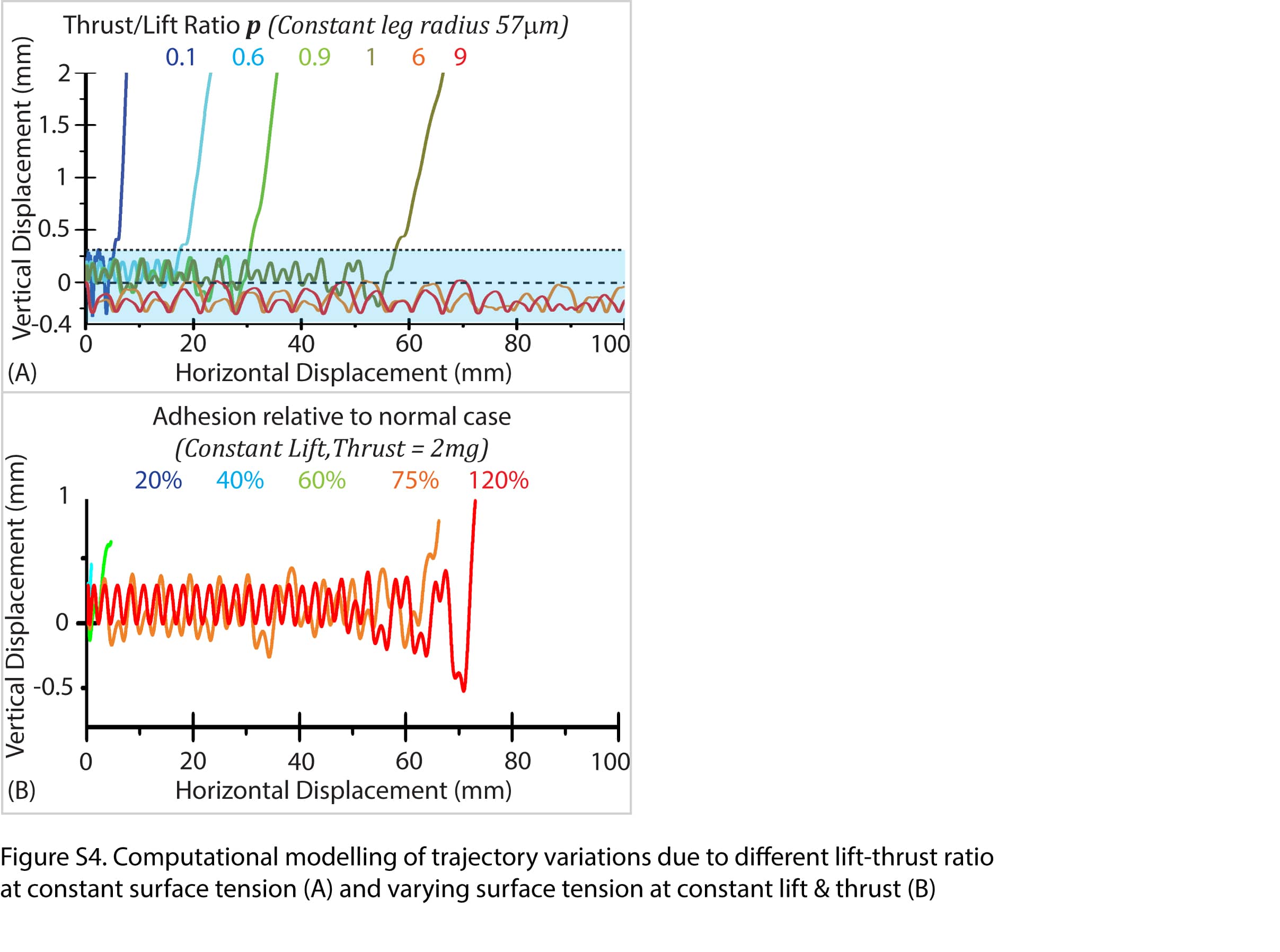}}
\end{figure}

\newpage
\noindent\textbf{Supplementary text and mathematical modeling}\\\\
\noindent\textendash\space Analytical model for forces acting during interfacial flight\\
\textendash\space Drag force calculations\\
\textendash\space Lift coefficient ($C_{L}$) estimation\\
\textendash\space Estimation of Lyapunov exponents\\

\noindent\textbf{Section 1 - Analytical reduced-order model for interfacial flight} \\
The forces acting on an insect in interfacial flight arise from four physical phenomena –
(1) Capillarity – capillary wave drag $C_{x}$ (horizontal) and surface tension $S_{y}$ (vertical)
(2) Aerodynamics – thrust $T_{x}$ (horizontal), lift $L_{y}$ (vertical) and air drag $A_{x}$ (horizontal)
(3) Bulk hydrodynamics – water drag on immersed legs $W_{x}$, $W_{y}$ (horizontal and vertical)
(4) Gravity – body weight $G_{y}$ (vertical)
Newton’s Second Law of motion for this system is formulated in the horizontal ($x$) and vertical ($y$) axes as $m\ddot{x}=T_{x}-W_{x}-A_{x}-C_{x}$ and $m\ddot{y}=L_{y}-W_{y}-G_{y}-S_{y}$. The expressions for each of these forces are –
\begin{equation}\label{Cx}C_{x}\simeq4*\rho_{w}c_{min}^{2}\kappa^{-1}\sqrt{\frac{\dot{x}}{c_{min}}-1}*2R_{claw}*U(\dot{x}-c_{min})\end{equation}
\begin{equation}\label{Sy}S_{y}=\begin{Bmatrix} 4*2\pi\sigma R_{claw} & , & y<-H_{claw}\\
\{4*2\pi\sigma R_{claw}\textrm{sech}(\frac{y-H_{claw}}{R_{claw}})*\frac{-y}{\left|y\right|} & , & -H_{claw}\leq y\leq H_{claw}\\ 0 & , & y>H_{claw} \end{Bmatrix}\end{equation}
\begin{equation}\label{Ly}L_{y}=L\sin(2\pi ft+\phi)*[U(\sin(2\pi ft+\phi))+r\{1-U(\sin(2\pi ft+\phi))\}]\end{equation}
\begin{equation}\label{Tx}T_{x} = pL_{y}\end{equation}
\begin{equation}\label{Ax}A_{x}=\frac{1}{2}\rho_{air}C_{D_{body}}*\pi ab*\dot{x}^{2}\end{equation}
\begin{equation}\label{Wd}W_{d}=4*\frac{1}{2}\rho_{w}C_{D_{leg}}\pi R_{claw}^{2}\dot{d^{2}},\thickspace\thickspace\thickspace d=x,y\end{equation}
\begin{equation}\label{Gy}G_{y}=-g\end{equation}
Here we derive detailed expressions for each of these individual forces.\\\\
\textit{Capillary wave drag}\\
The portion of the legs immersed just below the surface move through the interface at speeds greater than capillary wave speed to give rise to this force. The legs have a characteristic radius of $57\thinspace\mu$m which is small compared to the capillary length $\kappa^{-1}=2.709$ mm, typical horizontal velocities ranging from $0.3$ to $0.5$ ms$^{-1}$, which is comparable to the capillary wave speed of $0.23$ ms$^{-1}$, and Weber numbers of the order of $0.1$ are quite low. Hence, the dipolar approximation \cite{chepelianskii} at low Weber number, can be used to compute the capillary wave drag force on each leg. Using the force normalized per unit transverse extent of the submerged object, we multiply this by the leg diameter to get the total force. We obtain the expression in Eqn \eqref{Cx} \begin{equation*}C_{x}\simeq4*\rho_{w}c_{min}^{2}\kappa^{-1}\sqrt{\frac{\dot{x}}{c_{min}}-1}*2R_{claw}*U(\dot{x}-c_{min})\end{equation*}\\\\
\textit{Surface tension}\\
The surface tension on each leg is modelled assuming quasi-static deformation of non-interacting minimal surface menisci. These approximations are justified as the capillary relaxation time scale ($\approx 50\thinspace\mu$s) is much smaller than the inertial timescale (wingbeat period $\tau=8.67$ ms), meaning that any flows within the meniscus or contact line hysteresis die out fast enough that they can be ignored and meniscus deformations assumed to be instantaneous. Also, the meniscus maximum height ($\approx 250\thinspace\mu$m) is much smaller than the capillary length $\kappa^{-1}=2.709$ mm, which in turn is smaller than the physical separation between the legs corresponding to the beetle's body width and length of $4$ to $6$ mm, indicating that the menisci are minimal surfaces placed far enough apart that their finite physical size is too small for them to interact.  The four legs immersed in water are modelled as non-interacting infinite fibres, each of which forms a minimal surface meniscus with the contact line pinned at the line barrier between the superhydrophobic tarsus and the hydrophilic claw \cite{degennes}. The meniscus is described as the minimal curve \begin{equation}x=R_{claw}\textrm{cosh}(\frac{y-H_{claw}}{R_{claw}})\end{equation}
This minimal curve assumption was experimentally verified by numerically fitting the meniscus produced by dipping the claw of a dead beetle in water and pulling it up to create a meniscus of known vertical height (Figure 2K). The weight of water in the meniscus is supported by gravity, and hence the maximum height where the meniscus becomes too heavy and breaks is \begin{equation}H_{claw}=R_{claw}\ln(\frac{2\kappa^{-1}}{R_{claw}})\end{equation} when its horizontal extent reaches the capillary length.\\
There are three different cases possible for surface tension. First, when the leg is immersed deeper than the tarsal-claw joint, we assume contact angle is always $0^{\circ}$ and the contact line slips along the tarsus. Second, when the leg rises above the maximum vertical extent of the meniscus, the meniscus breaks releasing the leg from contact with water and there is nor more surface tension, unless the leg descends sufficiently to become re-immersed in water. Third, in the normally observed case, the leg has an intermediate displacement with contact angle being a function of displacement. Here, the contact angle $\theta$ at the fibre is calculated from the slope of the meniscus as \begin{equation}\cos(\theta)=\textrm{sech}(\frac{y-H_{claw}}{R_{claw}})\end{equation}
Taking these three regimes into account, the resultant force from surface tension \cite{degennes} is expressed in Eqn \eqref{Sy} as \begin{equation*}S_{y}=\begin{Bmatrix} 4*2\pi\sigma R_{claw} & , & y<-H_{claw}\\
\{4*2\pi\sigma R_{claw}\textrm{sech}(\frac{y-H_{claw}}{R_{claw}})*\frac{-y}{\left|y\right|} & , & -H_{claw}\leq y\leq H_{claw}\\ 0 & , & y>H_{claw} \end{Bmatrix}\end{equation*}
To represent the transition from 2D flight to airborne flight, we define a take-off time $T_{to}$, which is the last instant during a trajectory where the claw is in contact with water.\\\\
\textit{Thrust and Lift}\\
For the wing force, the contribution of ground effect \cite{rayner} due to interaction with the deformable water surface is unknown, and the exact variation of force with stroke position is also unknown for this system. Hence we loosely approximate wing force as a sinusoid \cite{dudley} with a fixed wingbeat frequency $f = 116 Hz$, as the kinematics of the wing tip follows a sinusoidal trajectory. We parameterize this expression with variables for amplitude of lift force $L$, ratio of thrust to lift $p$, wing stroke angle at initiation of motion $\phi$ and ratio of forces produced in upstroke and downstroke $r$ - as all of these can vary for each trajectory at the will of the insect, and cannot be easily measured without perturbing the insect. The total wing force is resolved into the two components of vertical lift and horizontal thrust in Eqns \eqref{Ly}, \eqref{Tx} \textendash
\begin{equation*}L_{y}=L\sin(2\pi ft+\phi)*[U(\sin(2\pi ft+\phi))+r\{1-U(\sin(2\pi ft+\phi))\}]\end{equation*}\begin{equation*}T_{x} = pL_{y}\end{equation*}
Here, $U$ is the Heaviside step function.\\\\
\textit{Air drag}\\
To compute the aerodynamic drags, the beetle's body is approximated as an ellipsoidal bluff body in cross-flow, with the flow incident on the ventral surface of the insect’s body. Air drag on the body is given in Eqn \eqref{Ax} by \begin{equation*}A_{x}=\frac{1}{2}\rho_{air}C_{D_{body}}*\pi ab*\dot{x}^{2}\end{equation*}
where $a=6$ mm and $b=4$ mm are the major and minor axes of the ellipse formed by the body cross-section and $C_{D_{body}}\sim1.5$ is a typical body drag coefficient for insects of similar size and profile \cite{nachtigall}.\\\\
\textit{Water drags}\\
For drag from water, we roughly approximate the pair of curved claws as a sphere, whose radius is assumed to be equal to the radius of the leg at the claw joint, $R_{claw}=57\thinspace\mu$m. Drag forces from water are expressed for spheres in cross-flow in Eqn \eqref{Wd} as \begin{equation*}W_{d}=4*\frac{1}{2}\rho_{w}C_{D_{leg}}\pi R_{claw}^{2}\dot{d^{2}}\end{equation*} where $C_{D_{leg}}\sim3$ is the drag coefficient for a sphere in cross-flow at the appropriate Reynolds number range $10<Re<100$, and $\dot{d}=\dot{x},\dot{y}$ respectively for horizontal and vertical drag forces.\\\\
\textit{Gravity}\\
The body weight is given by $mg$, where $m\approx2.2$ mg is the average weight of the insect calculated by averaging weights for about 30 dead insects. $g=9.81$ ms$^{-2}$ is the acceleration due to gravity.\\\\
\noindent\textit{Validation}\\
Simulations of the dynamics were performed in MATLAB R2012b using a built-in ODE solver (ode113), with relative tolerances of $10^{-9}$ and constant integration time step of $10^{-8}$ seconds. The elimination of numerical noise was verified by computing take-off times for a variety of different thrust-to-lift ratios $p$ keeping other parameters constant, and checking that take-off time is repeatably constant for a given lift $L$ at the required precision, as expected from decoupling of horizontal and vertical dynamics.\\
In our experiments on live \textit{G. nymphaeae}, there are natural variations in initial displacements and velocities $\{x_{o}, \dot{x_{o}}, y_{o}, \dot{y_{o}}\}$,  wingbeat frequency $f$, magnitude of wing lift $L$, thrust-to-lift ratio $p$, initial wing stroke phase $\phi$, and ratio of forces generated in upstroke and downstroke $r$. These $9$ parameters can even vary within a single trajectory, typically in the initial stages of flight when the insect has yet to settle into a consistent rhythm. Moreover, many of these, such as $L$, $p$, $\phi$ and $r$, are not directly measurable quantities. Since this non-linear system is expected to be sensitive to even small variations in these parameters, it is not possible to sweep across all of them to exactly fit each experimentally obtained trajectory. Hence, we validated our model by ensuring that simulated trajectories capture four basic kinematic features common to all experimental trajectories -\\
(1) Horizontal displacement has sigmoidal variations in each wingbeat superimposed on a quadratically increasing profile, covering between $15$ mm and $25$ mm in about $100$ ms.\\
(2) Horizontal velocity varies sinusoidally in each wingbeat, with the mean rising linearly to $0.25$ ms$^{-1}$ to $0.5$ ms$^{-1}$ before levelling off.\\
(3) Vertical displacement should show oscillations of amplitude $0.2$ and $0.5$ mm and frequency approximately equal to wingbeat.\\
(4) The peak displacement in each oscillation should vary in height.\\
Our model was able to reproduce simulated trajectories with these characteristics for physiologically feasible ranges of the parameters, as shown in Figure S3. Thus, we validate that our model is a good approximation that captures the most basic and essential physics involved in interfacial flight.\\\\

\noindent\textbf{Section 2 – Drag force calculations}\\
Consider a typical flight speed of the insect around $0.3$ ms$^{-1}$.  We refer to equations for drag forces derived Section 1. We assume that the insect has similar profile when flying in air or on the interface, with a comparable cross-sectional body area. For an insect flying in air, air drag estimated using Eqn \eqref{Ax} is around $ A_{x}=\frac{1}{2}\rho_{air}C_{D_{body}}*\pi ab*\dot{x}^{2}\approx6\thinspace\mu$N, using values of $a=6$ mm, $b=4$ mm and $C_{D_{body}}\sim1.5$. For an insect flying on the interface, the same air drag is experienced. In addition, there is capillary gravity wave drag, which is estimated from Eqn \eqref{Cx} as $C_{x}\simeq4*\rho_{w}c_{min}^{2}\kappa^{-1}\sqrt{\frac{\dot{x}}{c_{min}}-1}*2R_{claw}*U(\dot{x}-c_{min})\approx36\thinspace\mu$N. Hence, the total drag in interfacial flight is $42\thinspace\mu$N, which is about $7$ times the drag in airborne flight.
Air drag increases as the square of the velocity, which is a faster rise than capillary drag which increases as the square root of the velocity. However, the two become approximately equal only at $\dot{x}\approx 1.52$ ms$^{-1}$, which is an extremely high speed for interfacial motion!\\\\

\noindent\textbf{Section 3 - Lift coefficient ($C_{L}$) estimation} \\
We assume that the force generated by the wings during stalled vertical take-off is entirely directed into vertical lift, and that the contact angle on the two wetted legs is $180^{\circ}$. Since the insect is in static equilibrium, we estimate the lift using Eqn (3) to be $L_{max}=2*(2\pi \sigma R_{claw})+mg=73.2\thinspace\mu$N.
Parameters of the wing stroke are measured from Movie SV5 having a resolution of $115\thinspace\mu$m. The wing chord is measured to be $C=5.6$ mm. We measured the typical stroke angle as $\varphi =120^{\circ}$ using Movie SV5, which has a rear view. For a wingbeat frequency of $\nu=3000/26$ Hz, this corresponds to a wing angular velocity of $\omega =2\varphi \nu =483$ rad s$^{-1}$. Wingtip velocity is thus $V_{w}=C\omega =2.7$ ms$^{-1}$.
Drawing an free-hand tool outline around the face-on view of the wing in midstroke gives a wing area $S=11$ mm$^{2}$, measured using ImageJ. For verification, measuring wing chord and the broadest part of the wing in the vertical diection to be $5.6$ mm and $2.8$ mm respectively and assuming an elliptical wing gives an area of $\frac{\pi}{4} Cw\approx12$ mm$^{2}$, which is close to the measured value.
Hence, we can compute an approximate lift coefficient for the wing $C_{L}\simeq L_{max}/0.5\rho_{air}SV_{w}^{2}=1.55$.\\\\\\

\noindent\textbf{Section 4 – Characteristics of Lyapunov exponents} \\
Consider the dynamic equation for vertical motion for the vector $\overrightarrow{Y}=\{t,y,\dot{y}\}$ in phase space, for the case where the beetle is continuously attached to the interface, i.e. $y\leq H_{claw},\thickspace\forall\thickspace t\geq 0$. The second order inhomogenous ODE can be written as a set of 1st order ODEs as \begin{equation}\label{2orderode}\frac{d\overrightarrow{Y}}{dt}=\begin{bmatrix} \begin{smallmatrix}1\\ \dot{y}\\ \ddot{y}\end{smallmatrix} \end{bmatrix}=\begin{bmatrix} \begin{smallmatrix}1\\ \dot{y}\\ \frac{L}{m}\sin(2\pi ft+\phi)[U(sin(2\pi ft+\phi))+r(1-U(sin(2\pi ft+\phi)))]-\frac{2\rho_{w}C_{D_{leg}}\pi R_{claw}^{2}}{m}\dot{y}^{2}\textrm{sgn}(\dot{y})-\frac{8\pi \sigma R_{claw}}{m}\textrm{sech}(\frac{y-H_{claw}}{R_{claw}})\textrm{sgn}(y)\end{smallmatrix} \end{bmatrix}\end{equation}
For any small perturbation $\overrightarrow{\varepsilon}$, $\frac{d\overrightarrow{\varepsilon}}{dt}=\mathbb{J}\centerdot \overrightarrow{\varepsilon}$,  where $\mathbb{J}$ is the Jacobian matrix for the system, given by
\begin{equation}\label{jacobian}\mathbb{J}=\begin{bmatrix} \begin{smallmatrix} 0&0&0 \\ 0&0&1 \\ \{\frac{L}{m}\cos(2\pi ft+\phi)[U(sin(2\pi ft+\phi))+r(1-U(sin(2\pi ft+\phi)))]\}&\{\frac{8\pi \sigma R_{claw}}{m}\textrm{sech}(\frac{y-H_{claw}}{R_{claw}})\tanh(\frac{y-H_{claw}}{R_{claw}})\textrm{sgn}(y)\}&\{\frac{-4\rho_{w}C_{D_{leg}}\pi R_{claw}^{2}}{m}\lvert\dot{y}\rvert\}\end{smallmatrix} \end{bmatrix}\end{equation}
The eigenvalues of $\mathbb{J}$ are $\lambda_{1}=0, \lambda_{2,3}=-\mathbb{A} \pm \sqrt{\mathbb{A}^2+\mathbb{B}}$ where $\mathbb{A}=\frac{2\pi \rho_{w}C_{D_{leg}}R_{claw}^{2}}{m}\lvert\dot{y}\rvert$ and \\ $\mathbb{B}=\frac{8\pi \sigma R_{claw}}{m}\textrm{sech}(\frac{y-H_{claw}}{R_{claw}})\textrm{tanh}(\frac{y-H_{claw}}{R_{claw}})\textrm{sgn}(y)$. \\
$\lambda_{2,3} \thinspace\epsilon\thinspace \mathbb{R}$ for values of $\mathbb{A}^{2}+\mathbb{B}\geq 0$ and $(\lambda_{2}*\lambda_{3})<0$ for $\sqrt{\mathbb{A}^{2}+\mathbb{B}}>\mathbb{A}$. \\
\begin{equation}\implies \lambda_{2,3} \phantom{X}\mathbb{\epsilon}\phantom{X} \mathbb{R} \phantom{X}\&\phantom{X} (\lambda_{2}*\lambda_{3})<0 \phantom{X}\forall\phantom{X} \mathbb{B}>0\end{equation}
\begin{equation}\implies \textrm{sech}(\frac{y-H_{claw}}{R_{claw}})\textrm{tanh}(\frac{y-H_{claw}}{R_{claw}})\textrm{sgn}(y) >0\end{equation}\\
In the domain $y\leq H_{claw}$, $\textrm{sech}(\frac{y-H_{claw}}{R_{claw}})>0$ and $\textrm{tanh}(\frac{y-H_{claw}}{R_{claw}})<0$ always, $\implies \textrm{sgn}(y) < 0$ \\
$\therefore$ Eigenvalues of $\mathbb{J}$ are real and of opposite signs for $-H_{claw}\leq y<0$. \\
The transformation matrix for this system is $\Phi (t)=\exp{\int\limits_{0}^{t}\mathbb{J}dt}$. \\
The Lyapunov exponents of the system are hence equal to the eigenvalues $\lambda_{i}$ of $\mathbb{J}$ \\ $det\lvert\Phi (t)\rvert=det\lvert\exp(\int\limits_{0}^{t}\mathbb{J}dt)\rvert=\exp(\sum\limits_{i=1}^{3}\lambda_{i}t)$. \\ $\implies \lim_{t\rightarrow \infty}\frac{1}{t}\ln(det\lvert \Phi (t)\rvert)=-2\mathbb{A}$ which is finite and exists for all $t$.\\
$\lim_{t\rightarrow \infty}\frac{1}{t}\ln(det\lvert \Phi (t)\rvert)=\sum\limits_{i=1}^{3}\lambda_{i}$ for the $\mathbb{R}^{3}$ basis $\begin{bmatrix}\begin{smallmatrix}1\\0\\0\end{smallmatrix}\end{bmatrix}$, $\begin{bmatrix}\begin{smallmatrix}0\\1\\0\end{smallmatrix}\end{bmatrix}$, $\begin{bmatrix}\begin{smallmatrix}0\\0\\1\end{smallmatrix}\end{bmatrix}$. \\\\
Hence, the vertical dynamics has a transformation matrix $\Phi (t)$ which is regular, implying that the Lyapunov exponents $\lambda_{i}, i = 1,2,3$ exist and are finite for all perturbations of trajectories and are independent of the initial conditions.\\ In the domain $-H_{claw} \leq y \leq H_{claw}$, $\sum\limits_{i=1}^3 \lambda_{i} = -2\mathbb{A} < 0$ always, hence the system is always dissipative, as expected of motion under drag forces. Additionally, when $-H_{claw} \leq y<0$, the exponents $\lambda_{2}, \lambda_{3}$ are real and of opposite signs. Hence those regimes of a trajectory where $-H_{claw} \leq y<0$ evolve along chaotic attractors. Therefore, every trajectory that lies at least partially in a region of phase space where these two conditions hold true displays chaotic behaviour. The downward pull of gravity on the beetle results in the equilibrium vertical displacement being below the mean water level at infinity ($y_{eq}<0$). Therefore at low values of lift force $L$, the resulting oscillatory trajectories confined to the interface typically have $y<0$ for significant continuous portions of the trajectory, resulting in chaotic motion.\\

\newpage
\noindent\textbf{SUPPLEMENTARY VIDEOS}\\
(Available online at https://goo.gl/CcCfkt)\\\\
\textbf{Video SV1: Side view close-up of interfacial flight in \textit{Galerucella nymphaeae}}\\
Side view close-up video of interfacial flight, used for trajectory extraction. The hind leg femur-tibia joint is visible as a dark spot, and was tracked across images. Note the initial lifting of the middle legs prior to flight. Recorded at 3000 fps, playback at 15 fps. Beetle body length is 6 mm.\\\\
\textbf{Video SV2: Side view of interfacial flight}\\
Side view close-up of initial posture adjustment and initiation of interfacial flight, showing postural changes and highlighting the role of the immersed claw in contact line pinning. Recorded at 3000 fps, playback at 30 fps. Beetle body length is 6 mm.\\\\
\textbf{Video SV3: Front view of takeoff into air}\\
Front view of G.nymphaeae taking off from the water surface. Recorded at 3000 fps, playback at 20 fps. Beetle body length is 6 mm.\\\\
\textbf{Video SV4: Rear view of takeoff into air}\\
Rear view of G.nymphaeae taking off from the water surface. Recorded at 3000 fps, playback at 20 fps. Beetle body width is 4 mm.\\\\
\textbf{Video SV5: Rear view close-up of takeoff into air}\\
Rear view close-up of initial posture adjustment and initiation of interfacial flight. Recorded at 3000 fps, playback at 30 fps. Beetle body width is 4 mm.\\\\
\textbf{Video SV6: Capillary wave formation}\\
Formation of capillary waves on the surface of water during interfacial flight, visible as distortions of the edges of the water dish and optical posts supporting the experimental setup. Recorded at 3000 fps, playback at 30 fps. Beetle body length is 6 mm.\\\\
\textbf{Video SV7: Seamless interfacial flight and take-off into air}\\
Smooth variation between interfacial and airborne flight in a single flight sequence. Recorded at 3000 fps, playback at 30 fps. Beetle body length is 6 mm.\\\\
\textbf{Video SV8: Backward Flight}\\
Backward movement of beetle in air on taking off from a pedestal. Recorded at 1000 fps, playback at 20 fps. Beetle body length is 6 mm.\\\\
\textbf{Video SV9: Failed take-off flight}\\
Failed take-off attempt where take-off failed due to wetting of the beetle's hind legs. Initial detachment of middle legs from the surface and postural tilt of the beetle's body axis from horizontal to vertical are clearly seen, although pixel resolution is insufficient to accurately measure meniscus deformations. Recorded at 3000 fps, playback at 30 fps. Beetle body length is 6 mm.\\\\
\textbf{Video SV10: Flight tumbling}\\
Tumbling of beetle during 2D flight on a dish and subsequent recovery of upright posture. Recorded at 30 fps, playback at 30 fps (real time). Beetle body length is 6 mm.\\\\
\textbf{Video SV11: Lack of braking mechanism in interfacial flight}\\
Interfacial flight sequence in G.nymphaeae on the surface of water in a Petri dish, where the insect does not stop but falls off the edge. Recorded at 3000 fps, playback at 30 fps. Beetle body length is 6 mm.\\\\
\textbf{Video SV12: Breaking of meniscus during takeoff}\\
Formation of meniscus at the hind legs, which is lifted up to maximum height and subsequently breaks contact during takeoff, setting up small ripples on the interface. Recorded at 3000 fps, playback at 30 fps. Beetle body length is 6 mm.

\end{document}